\documentclass[reqno,11pt]{amsart}
\usepackage{graphicx}
\usepackage{slashed}
\usepackage{amscd}
\usepackage{tensor}
\usepackage{amssymb}
\usepackage{esint}
\usepackage{enumitem}
\usepackage{accents}
\usepackage[usenames,dvipsnames]{pstricks}
\usepackage{pst-grad} 
\usepackage{pst-plot} 
\usepackage[mathscr]{eucal}
\numberwithin{equation}{section}
\allowdisplaybreaks[1]

\title[Perturbation Theory for Causal Variational Principles]{Perturbation Theory
for Critical Points of \\ Causal Variational Principles}

\author[F.\ Finster]{Felix Finster \\ \\ March 2017 / December 2018}
\address{Fakult\"at f\"ur Mathematik \\ Universit\"at Regensburg \\ D-93040 Regensburg \\ Germany}
\email{finster@ur.de}

\newtheorem{Def}{Definition}[section]

\newtheorem{Prp}[Def]{Proposition}
\newtheorem{Lemma}[Def]{Lemma}
\newtheorem{Remark}[Def]{Remark}

\newtheorem{Example}[Def]{Example}

\newcommand{\Thanks}{\vspace*{.5em} \noindent \thanks}
\newcommand{\beq}{\begin{equation}}
\newcommand{\eeq}{\end{equation}}
\newcommand{\Proof}{\begin{proof}}
\newcommand{\QED}{\end{proof} \noindent}
\newcommand{\QEDrem}{\ \hfill $\Diamond$}

\newcommand{\la}{\langle}
\newcommand{\ra}{\rangle}

\newcommand{\Sl}{\mbox{$\prec \!\!$ \nolinebreak}}
\newcommand{\Sr}{\mbox{\nolinebreak $\succ$}}

\newcommand{\C}{\mathbb{C}}
\newcommand{\R}{\mathbb{R}}
\newcommand{\1}{\mbox{\rm 1 \hspace{-1.05 em} 1}}

\newcommand{\N}{\mathbb{N}}
\newcommand{\Pdd}{\mbox{$\partial$ \hspace{-1.2 em} $/$}}

\DeclareMathOperator{\Tr}{Tr}
\DeclareMathOperator{\tr}{tr}
\renewcommand{\O}{{\mathscr{O}}}
\renewcommand{\L}{{\mathcal{L}}}
\newcommand{\Sact}{{\mathcal{S}}}
\newcommand{\T}{{\mathcal{T}}}

\newcommand\B{{\mathscr{B}}}
\newcommand{\U}{\text{\rm{U}}}

\newcommand{\g}{{\mathfrak{g}}}

\DeclareMathOperator{\supp}{supp}
\renewcommand{\H}{\mathscr{H}}

\newcommand{\Lin}{\text{\rm{L}}}
\newcommand{\F}{{\mathscr{F}}}

\DeclareMathOperator{\re}{Re}

\newcommand{\bitem}{\begin{itemize}[leftmargin=2.5em]}
\newcommand{\eitem}{\end{itemize}}
\newcommand{\itemD}{\item[{\raisebox{0.125em}{\tiny $\blacktriangleright$}}]}

\DeclareFontFamily{OT1}{rsfso}{}
\DeclareFontShape{OT1}{rsfso}{m}{n}{ <-7> rsfso5 <7-10> rsfso7 <10-> rsfso10}{}
\DeclareMathAlphabet{\mycal}{OT1}{rsfso}{m}{n}

\newcommand{\s}{\mathfrak{s}}
\newcommand{\J}{\mathfrak{J}}
\newcommand{\Jdiff}{\mathfrak{J}^\text{\rm{\tiny{diff}}}}
\newcommand{\Jtest}{\mathfrak{J}^\text{\rm{\tiny{test}}}}
\newcommand{\Jlin}{\mathfrak{J}^\text{\rm{\tiny{lin}}}}
\newcommand{\Jc}{\mathfrak{J}^\text{\rm{\tiny{c}}}}
\newcommand{\lin}{\text{\rm{\tiny{lin}}}}
\newcommand{\comp}{\text{\rm{\tiny{c}}}}
\newcommand{\Jfermi}{\mathfrak{J}^\text{\rm{\tiny{f}}}}
\newcommand{\Jbose}{\mathfrak{J}^\text{\rm{\tiny{b}}}}

\newcommand{\bJtest}{\bar{\mathfrak{J}}^\text{\rm{\tiny{test}}}}

\newcommand{\Fluct}{\text{\rm{\tiny{F}}}}
\renewcommand{\u}{\mathfrak{u}}
\renewcommand{\v}{\mathfrak{v}}
\newcommand{\w}{\mathfrak{w}}
\newcommand{\Ctest}{C^\text{\rm{\tiny{test}}}}
\newcommand{\Gdiff}{\Gamma^\text{\rm{\tiny{diff}}}}
\newcommand{\Gtest}{\Gamma^\text{\rm{\tiny{test}}}}

\newcommand{\Gfermi}{\Gamma^\text{\rm{\tiny{f}}}}
\newcommand{\Gbose}{\Gamma^\text{\rm{\tiny{b}}}}
\newcommand{\G}{\mathscr{G}}

\renewcommand{\u}{\mathfrak{u}}
\newcommand{\as}{{\mathfrak{a}}}
\newcommand{\bs}{{\mathfrak{b}}}

\begin{document}

\maketitle

\vspace*{-0.8cm}

\begin{abstract}
The perturbation theory for critical points of causal variational principles is developed.
We first analyze the class of perturbations obtained by multiplying the universal measure by a weight
function and taking the push-forward under a diffeo\-mor\-phism.
Then the constructions are extended to convex combinations of such measures,
leading to perturbation expansions for the mean and the fluctuation of the measure,
both being coupled in higher order perturbation theory.
It is explained how our methods and results apply to the causal action principle for causal fermion systems.
It is shown how the perturbation expansion in the continuum limit and the effect of
microscopic mixing are recovered in specific limiting cases.
\end{abstract}

\tableofcontents

\section{Introduction}
The theory of causal fermion systems is a recent approach to fundamental physics. Giving
quantum mechanics, general relativity and quantum field theory as limiting cases, it is a candidate for a unified
physical theory (see~\cite{cfs} or the survey article~\cite{dice2014}).
So far, the connection to perturbative quantum field theory has been established
by first taking the continuum limit~\cite{cfs}
and then including the mechanism of microscopic mixing (see~\cite{qft}).
Although this procedure gives the correct limiting case with an interaction described by
a unitary time evolution on Fock spaces (see~\cite[Section~8]{qft}),
the derivation is not quite convincing conceptually because it is based on the
perturbation expansion for solutions of the Dirac equation coupled to classical bosonic fields
as obtained in the continuum limit (see~\cite[\S3.8.4]{cfs} and~\cite[Section~2]{qft}).
In order to clarify the mathematical structure of the theory, it is desirable to perform the perturbation
expansion directly for the universal measure of the causal fermion system, without
referring to specific limiting cases (for more details see Section~\ref{secmotivate} below).
Analyzing this problem also opens up the research program to explore
how the perturbation theory for causal fermion systems differs from perturbative quantum field
theory, with the goal of making experimental predictions.

In this paper the general perturbation theory for causal fermion systems is developed.
We thus succeed in extending the methods of perturbative quantum field theory
to non-smooth situations where space-time has a non-trivial, possibly discrete
microscopic structure and the physical equations are no longer obtained
by quantizing differential equations.
We work in the jet formalism introduced in~\cite{jet} in the more general and at the same time more
convenient framework of causal variational principles in the non-compact setting.
Our perturbation expansion has the nice feature that the bosonic and fermionic
perturbations are described on the same footing in terms of jet spaces containing bosonic
and fermionic subspaces.

In the setting of causal variational principles, the basic object is a measure~$\rho$
on a manifold~$\F$ (for the necessary preliminaries see Section~\ref{secprelim}).
Our methods for perturbing this measure are developed in two steps.
In the first step, our method is to multiply~$\rho$ by a non-negative function~$f$ and to
take the push-forward under a mapping~$F$,
\beq \label{nofragment}
\tilde{\rho} = F_* \big( f \, \rho \big) \:.
\eeq
We then compute~$f$ and~$F$ order by order in a formal power expansion in a
``coupling parameter''~$\lambda$.
In the second step, we consider more generally 
a convex combination of a finite number of measures of the form~\eqref{nofragment},
\beq \label{withfragment}
\tilde{\rho} = \frac{1}{L} \sum_{\as=1}^L (F_\as)_* \big( f_\as \, \rho \big) \:.
\eeq
This ansatz allows for the possibility that the
measure is ``decomposed'' into several components and the 
support of the measure is ``enlarged'' (see Figure~\ref{figfragment}
on page~\pageref{figfragment}).
We refer to this effect as a {\em{fragmentation}} of the measure.
In analogy to the perturbation theory for degenerate eigenvalues of a linear operator,
where the perturbation must be ``diagonalized on the degenerate eigenspace'' before
performing the perturbation expansion, the perturbation theory with fragmentation
makes it necessary to choose jets which describe how the fragmentation
forms (see~\eqref{f0choice} and~\eqref{F1choice} on page~\pageref{f0choice}).

The paper is organized as follows. In Section~\ref{secmotivate} we give a brief
physical motivation and put our perturbation expansion into the 
context of the ongoing research program on causal fermion systems.
Section~\ref{secprelim} provides the necessary
background on causal variational principles and the jet formalism.
In Section~\ref{secabstract} the perturbation theory without fragmentation is developed.
After bringing the combinatorics into a convenient form (Section~\ref{secrhopert}),
we invert the linearized equations with Green's operators
(see Definition~\ref{defS} in Section~\ref{secgreen}). The resulting perturbation expansion is
summarized in Section~\ref{secdiagram}.
In Section~\ref{seclinpert} it is explained how, starting from a linearized solution,
one can construct a one-parameter family of nonlinear solutions of the field equations.
In Sections~\ref{secvacpert} it is shown how, perturbing the vacuum by an inhomogeneity,
one can construct a corresponding nonlinear solution of the field equations.

In Section~\ref{secfragment} the perturbation theory with fragmentation is developed.
The method is to decompose suitable jets describing the perturbation into their mean and 
the fluctuations (see~\eqref{decompose} in Section~\ref{seclinfluct}).
A technical complication is that, if fragmentation occurs, the unperturbed Laplacian
can no longer be inverted. This is illustrated in Section~\ref{secexample}
in a concrete example. The method to overcome this problem is to invert instead
the perturbed Laplacian (see Section~\ref{secpertfluct}).

In Section~\ref{seccfs} we explain how our methods and results apply to the setting of causal fermion systems.
After the necessary preliminaries (Section~\ref{seccfsprelim}), the
perturbation expansion for the wave evaluation operator is derived (Section~\ref{secweo}).
After identifying jets with perturbations of the wave evaluation operator,
the general perturbation expansion applies in a straightforward way (Section~\ref{seccfsvacpert}).

In Section~\ref{seccontinuum}, it is shown that by a suitable choice of the jet spaces
one recovers the analysis in the continuum limit as carried out in~\cite{cfs}.
Finally, in Section~\ref{secmixing} we describe how to incorporate the effect of microscopic mixing
as analyzed in~\cite{qft}.

\section{Physical Motivation and Significance of the Perturbation Expansion} \label{secmotivate}
Before delving into the constructions, we give a physical motivation of the perturbation expansion
and explain its significance within the research program on causal fermion systems
and causal variational principles.

We begin with a brief introduction and an outline of the present status of the theory.
Causal fermion systems are based on a novel {\em{mathematical model of space-time}},
where the basic object is a measure on linear operators of a Hilbert space:
\begin{Def} \label{defparticle} (causal fermion system) {\em{ 
Given a separable complex Hilbert space~$\H$ with scalar product~$\la .|. \ra_\H$
and a parameter~$n \in \N$ (the {\em{``spin dimension''}}), we let~$\F \subset \Lin(\H)$ be the set of all
selfadjoint operators on~$\H$ of finite rank, which (counting multiplicities) have
at most~$n$ positive and at most~$n$ negative eigenvalues. On~$\F$ we are given
a positive measure~$\rho$ (defined on a $\sigma$-algebra of subsets of~$\F$), the so-called
{\em{universal measure}}. We refer to~$(\H, \F, \rho)$ as a {\em{causal fermion system}}.
}}
\end{Def} \noindent
This definition gives rise to a space-time together with structures therein,
most notably a causal structure, spinorial wave functions and geometric objects like connection
and curvature. The resulting abstract setting has been worked out in a satisfying way
(see for example~\cite[Section~1.1]{cfs}).
Moreover, it is clear how the abstract structures are related to the usual objects in Minkowski space
or on a Lorentzian manifold (see~\cite[Section~1.2]{cfs} or the introduction and survey in~\cite{nrstg}).
In order to see the correspondence, one must keep in mind that
the objects of the causal fermion system involve an ultraviolet regularization
on a length scale~$\varepsilon>0$. Thus we always consider the regularized quantities as those
having mathematical and physical significance. 
The corresponding objects in Minkowski space
or on a Lorentzian manifold are obtained in a certain limiting case~$\varepsilon \searrow 0$
in which the ultraviolet regularization is removed.

In the theory of causal fermion systems, the physical equations are formulated
via a variational principle, the so-called {\em{causal action principle}}.
It is defined as follows (for more details see~[F1, \S1.1.1] or~\cite{jet}).
Given~$x, y \in \F$,  we denote the non-trivial eigenvalues of the operator product~$xy$
counting algebraic multiplicities by~$\lambda^{xy}_1, \ldots, \lambda^{xy}_{2n} \in \C$.
We introduce the {\em{spectral weight}}~$| \,.\, |$
of an operator as the sum of the absolute values
of its eigenvalues. In particular, the spectral weights of the operator
products~$xy$ and~$(xy)^2$ are defined by
\beq \label{sw}
|xy| = \sum_{i=1}^{2n} \big| \lambda^{xy}_i \big|
\qquad \text{and} \qquad \big| (xy)^2 \big| = \sum_{i=1}^{2n} \big| \lambda^{xy}_i \big|^2 \:.
\eeq
We introduce the Lagrangian and the causal action by
\begin{align}
\text{\em{Lagrangian:}} && \L(x,y) &= \big| (xy)^2 \big| - \frac{1}{2n}\: |xy|^2 \label{Lagrange} \\
\text{\em{causal action:}} && \Sact(\rho) &= \iint_{\F \times \F} \L(x,y)\: d\rho(x)\, d\rho(y) \:. 
\end{align}
The {\em{causal action principle}} is to minimize~$\Sact$ by varying the universal measure
under the following constraints,
\begin{align*}
\text{\em{volume constraint:}} && \rho(\F) = \text{const} \quad\;\; & \\
\text{\em{trace constraint:}} && \int_\F \tr(x)\: d\rho(x) = \text{const}&  \\
\text{\em{boundedness constraint:}} && \T(\rho) := \iint_{\F \times \F} |xy|^2\: d\rho(x)\, d\rho(y) &\leq C \:,
\end{align*}
where~$C$ is a given parameter (and~$\tr$ denotes the trace of a linear operator on~$\H$).
The form of the above Lagrangian is the result of
long considerations and many computations (for a systematic account see~\cite[Chapter~5]{pfp}).
The constraints are needed in order to obtain a mathematically well-defined variational
principle with non-trivial minimizers.
A simple way of understanding the structure of the Lagrangian is the following connection
to {\em{causality}}: Writing the Lagrangian as (see~\cite[eq.~(1.1.9)]{cfs})
\[ \L = \frac{1}{4n} \sum_{i,j=1}^{2n} \Big( \big|\lambda^{xy}_i \big| - \big|\lambda^{xy}_j \big| \Big)^2 \:, \]
one sees that~$\L$ vanishes if the eigenvalues~$\lambda^{xy}_i$ all have the
same absolute value. Defining spacelike separation by this property,
pairs of points with spacelike separation do not enter the action.
This can be seen in analogy to the usual notion of causality where
points with spacelike separation cannot influence each other.
It turns out that  in suitable limiting cases, the above definition of causality indeed agrees with the
usual causal structure of Minkowski space or of a Lorentzian manifold
(for details see~\cite[\S1.2.5]{cfs} or~\cite[Sections~4 and~5]{lqg}).

{\em{Causal variational principles}} are a mathematical generalization of the causal action principle,
with the aim of restricting attention to the essential analytic structures.
The existence theory and the general structure of the corresponding Euler-Lagrange
equations have been worked out in~\cite{continuum, lagrange}.
The connection to physics is made in~\cite[Chapter~3-5]{cfs}, where
it is shown that in a well-defined limiting case, the so-called {\em{continuum limit}},
the interaction given by the causal action principle can be described
effectively by the Dirac equation coupled to classical field equations for
gauge fields and the gravitational field.
In this limiting case, one obtains all the interactions of the standard model plus classical gravity.

The next challenge is to understand how {\em{quantum field theory}}
arises from the causal action principle. Indeed, in the paper~\cite{qft}
the connection between the causal action principle and a second-quantized dynamics on Fock spaces
has been made in a certain limiting case. But some of the assumptions and constructions
remain to be justified and understood better.
Moreover, a number of important open questions still need to be addressed:
\begin{itemize}[leftmargin=2em]
\item[(a)] Since an ultraviolet regularization on the scale~$\varepsilon$ is built in,
the theory of causal fermion systems is ultraviolet finite.
Nevertheless, it is an important task to understand the asymptotics of interacting systems
for small~$\varepsilon$. In particular, is it possible to take
limit~$\varepsilon \searrow 0$ with renormalization techniques?
Is the effective theory obtained in this limit {\em{renormalizable}}?
\item[(b)] The fact that the continuum limit also gives the Einstein equations
raises the question to which extent and how precisely the constructions in~\cite{qft}
relate to {\em{quantum gravity}}. Do causal fermion systems really give 
a mathematically well-defined setting for describing quantized gravitational fields?
Can the resulting ``geometry of quantum gravity'' be described by the geometric structures
of the causal fermion system?
\end{itemize}
Answering these questions in the affirmative would show that
causal fermion systems are a mathematically consistent, non-perturbative quantum
field theory. The ultimate goal is to understand the quantum field theory limit of causal
fermion systems in a way where it becomes possible to go beyond quantum field
theory in the following sense:
\begin{itemize}[leftmargin=2em]
\item[(c)] How does the dynamics described by the causal action principle
differ from quantum field theory? How can the deviations be quantified?
Can they be detected in experiments?
\end{itemize}

The present paper is an important step towards answering these questions, as we now
explain. The procedure in~\cite{qft} is closely tied to the continuum limit analysis
and to the Dirac equation coupled to classical field equations obtained in this limit.
The method to go beyond the continuum limit is referred to as {\em{microscopic mixing}}.
In the present formulation with measures, microscopic mixing can be understood
by taking universal measures~$\rho_1, \ldots, \rho_L$, each describing
a system of Dirac particles in Minkowski space with an interaction via classical
bosonic potentials~$A_1, \ldots, A_L$.
Then the convex combination of the measures
\beq \label{convex}
\tilde{\rho} = \frac{1}{L} \sum_{\as=1}^L \rho_\as
\eeq
is again a measure. It contains information on the bosonic potentials~$A_1, \ldots, A_L$
of the ``subsystems'' described by~$\rho_1, \ldots, \rho_L$.
As observed in~\cite{qft} (based on preliminary considerations in~\cite{entangle}),
the resulting collection of bosonic potentials can be described effectively
by a {\em{second-quantized}} bosonic field.
Moreover, taking into account an interaction of the subsystems described by~$\rho_1, \ldots, \rho_L$
by combining the perturbation expansion for classical fields with some features of microscopic mixing,
one obtains an effective interaction described by a second-quantized Hamiltonian acting on fermionic and bosonic
Fock spaces (see~\cite[Section~8]{qft}).

Although these constructions are an important first step,
there is the major shortcoming that the connection to Fock spaces is
based on the perturbation expansion for classical fields in each subsystem.
The fact that the description with classical fields is valid only approximately
makes it difficult to justify the validity and to quantify the error of the Fock space dynamics.
Moreover, the description makes it necessary to assume that there are subsystems,
but it remains unclear how the subsystems form dynamically.
In~\cite{qft}, this open problem was bypassed by considering the so-called
limiting case of an {\em{instantaneous recombination}} of subsystems.
But in order to tackle the above open questions,
the validity of this limiting case must be justified, 
and the errors of the approximation must be quantified.

The main point of the constructions of the present paper is to overcome the
shortcomings of the treatment in~\cite{qft}.
The perturbation expansion developed here only uses the Euler-Lagrange equations
of the causal action, but it does not rely on the classical field equations obtained
in the continuum limit. The possibility for the formation of subsystems~\eqref{convex}
is now taken into account by the fragmentation of the measure~\eqref{withfragment}.
In contrast to the ad-hoc ansatz~\eqref{convex}, the perturbation theory with fragmentation
makes it possible to analyze in detail whether and how fragmentation forms.
Moreover, the mutual interaction of the resulting subsystems can be quantified.
Intuitively speaking, the fragmentation of the measure means that space-time
does not stay classical, but becomes a ``quantum space-time'' which can be
thought of as a ``superposition'' or ``mixture'' of the space-times described by the
individual subsystems. In view of the general scope and applicability of our constructions,
the methods and results of this paper are a promising starting point for addressing
the above questions~(a)--(c) in a precise mathematical setting.

\section{Preliminaries} \label{secprelim}
\subsection{Causal Variational Principles in the Non-Compact Setting} \label{secnoncompact}
We consider causal variational principles in the non-compact setting
as introduced in~\cite[Section~2]{jet} (the connection to causal fermion systems
will be made in Section~\ref{seccfs} below). Thus let~$\F$ be a (possibly non-compact) smooth manifold
of dimension~$m \geq 1$.
Moreover, we are given a non-negative function~$\L : \F \times \F \rightarrow \R^+_0$
(the {\em{Lagrangian}}) with the following properties:
\begin{itemize}[leftmargin=2em] \label{assumotions1}
\item[(i)] $\L$ is lower semi-continuous, i.e.\ for all sequences~$x_n \rightarrow x$ and~$y_{n'} \rightarrow y$,
\[ \L(x,y) \leq \liminf_{n,n' \rightarrow \infty} \L(x_n, y_{n'})\:. \]
\item[(ii)] $\L$ is symmetric: $\L(x,y) = \L(y,x)$ for all~$x,y \in \F$.
\end{itemize}
Next, we let~$\rho$ be a (positive) Borel measure on~$\F$ (the {\em{universal measure}}).
The {\em{causal variational principle}} is to minimize the action
\[ \Sact = \int_\F d\rho(x) \int_\F d\rho(y)\: \L(x,y) \]
under variations of the measure~$\rho$, keeping the total volume~$\rho(\F)$ fixed.
If the total volume is infinite, one can make mathematical sense of variations of~$\Sact$
by considering variations of~$\rho$ of finite total variation and zero volume
(for details see~\cite[Section~2]{jet}). Here we do not enter the details of the
minimization procedure and of the properties of the minimizing measure.
Instead, we restrict attention to the resulting Euler-Lagrange (EL) equations
as derived in~\cite[Lemma~2.3]{jet}:
\begin{Def} \label{defcritical}
A Borel measure~$\rho$ on~$\F$ is a {\bf{minimizer}}
of the causal variational principle if it has the following properties:
\begin{itemize}[leftmargin=2em]
\itemD The measure~$\rho$ is locally finite.
\itemD The function~$\L(x,.)$ is $\rho$-integrable for all~$x \in \F$.
\itemD For a suitable value of the parameter~$\s>0$,
the function~$\ell$ defined by
\beq \label{ldef}
\ell(x) = \int_\F \L(x,y)\: d\rho(y) - \s
\eeq
is minimal and vanishes on the support of~$\rho$,
\beq \label{EL1}
\ell|_{\supp \rho} \equiv \inf_\F \ell = 0 \:.
\eeq
\end{itemize}
\end{Def}

We remark that the value of the parameter~$\s$ can be changed arbitrarily
by rescaling the measure according to
\beq \label{rescale1}
\rho \rightarrow \nu \rho \qquad \text{with} \qquad \nu>0 \:.
\eeq
With this in mind, we shall always keep~$\s$ fixed
when varying or perturbing the measure.

\subsection{The Weak Euler-Lagrange Equations}
Let~$\rho$ be a critical point of the causal variational principle.
We introduce {\em{space-time}}~$M$ as the support of this measure,
\[ M := \supp \rho \subset \F \:. \]
The idea behind the formulation of the weak EL equations is to use only part
of the information contained in the EL equations~\eqref{EL1}. Namely, we
evaluate them only on~$M$, taking into account first derivatives.
Moreover, we restrict attention to directions where~$\ell$ is differentiable.
And finally, we want to have the freedom to restrict attention to the part
of information needed for the application.
This leads to the following construction:
By~$C^\infty(M, \R)$ we denote all real-valued functions on~$M$ which have a smooth
extension to~$\F$. Likewise, by
\[ 
\Gamma = C^\infty(M, T\F) \]
we denote the smooth vector fields on~$\F$ along~$M$
(thus every $u \in \Gamma$ is a mapping from~$M$ to~$T\F$ with~$u(x) \in T_x\F$
for all~$x \in M$, which can be extended to a smooth vector field on~$\F$).
We define the jet space on~$M$ as the vector space
\[ 
\J := \big\{ \u = (a,u) \text{ with } a \in C^\infty(M, \R) \text{ and } u \in \Gamma \big\} \:. \]
Moreover, we let~$\Gdiff$ be those vector fields for which the
directional derivative of the function~$\ell$ exists,
\[ \Gdiff = \big\{ u \in C^\infty(M, T\F) \;\big|\; \text{$D_{u} \ell(x)$ exists for all~$x \in M$} \big\} \:. \]
Next, we introduce the subspace of jets
\[ \Jdiff := C^\infty(M, \R) \oplus \Gdiff \;\subset\; \J \:. \]
For a jet~$\u = (a, u) \in \Jdiff$ we define~$\nabla_\u$ as the linear combination of
scalar multiplication and directional derivative, i.e.
\[ \nabla_{\u} \ell(x) := a(x)\, \ell(x) + \big(D_u \ell \big)(x) \:. \]
Finally, we choose a linear subspace~$\Jtest \subset \Jdiff$ with the property
that its scalar and vector components are both vector spaces,
\[ \Jtest = \Ctest(M, \R) \oplus \Gtest \;\subset\; \Jdiff \:, \]
and the scalar component is nowhere trivial in the sense that
\[ 
\text{for all~$x \in M$ there is~$a \in \Ctest(M, \R)$ with~$a(x) \neq 0$}\:. \]
Then the {\em{weak EL equations}} read (for details cf.~\cite[(eq.~(4.10)]{jet})
\beq \label{ELtest}
\nabla_{\u} \ell|_M = 0 \qquad \text{for all~$\u \in \Jtest$}\:.
\eeq
The purpose of introducing~$\Jtest$ is that it gives the freedom to restrict attention to the portion of
information in the EL equations which is relevant for the application in mind.
For example, if one is interested only in the macroscopic dynamics, one can choose~$\Jtest$
to be composed of jets which disregard the microscopic fluctuations of~$\ell$.

We finally point out that the weak EL equations~\eqref{ELtest}
do not hold only for minimizers, but also for critical points of
the causal action. With this in mind, all methods and results of this paper do not apply only to
minimizers, but more generally to critical points of the causal variational principle.
For brevity, we also refer to a measure with satisfies the weak EL equations~\eqref{ELtest}
as a {\em{critical measure}}.

\subsection{Jet Spaces and the Linearized Field Equations}
For the detailed study of the weak EL equations it is most convenient
work with Taylor expansions of the component functions in a given chart.
Therefore, for any~$x \in M$ we choose a chart
of~$\F$ around~$x$  and work in components~$x^\alpha$.
For ease in notation, we usually omit the index~$\alpha$ as well as all vector and tensor indices.
But one should keep in mind that from now on, we always work in suitably chosen charts.

We now introduce useful jet spaces.
We begin with the space of {\em{dual jets}}~$(\Jtest)^*$. To this end, we denote the
continuous global one-jets taking values in the cotangent bundle restricted to~$M$ by
\[ \J^* := C^0(M, \R) \oplus C^0(M, T^*\F) \:. \]
We let~$(\Jtest)^*$ be the quotient space
\begin{align*}
(\Jtest)^* &:= \J^* \Big/  \big\{ (g,\varphi) \in \J^* \:\big|\: g(x) \,a(x) + \la \varphi(x), u(x) \ra = 0 \\
&\qquad\qquad\qquad\qquad\quad  \text{ for all~$\u=(a,u) \in \Jtest$ and all~$x \in M$} \big\} \:,
\end{align*}
where~$\la .,. \ra$ denotes the dual pairing of~$T^*_x\F$ and~$T_x\F$.
Here we take equivalence classes simply because it is
convenient to disregard dual jets which are trivial on~$\Jtest$.

We next introduce the spaces~$\J^\ell$, where the superscript~$\ell \in \N_0 \cup \{\infty\}$ can be
thought of as the order of differentiability if the derivatives act simultaneously on
both arguments of the Lagrangian:
\begin{Def} \label{defJvary}
For any~$\ell \in \N_0 \cup \{\infty\}$, the jet space~$\J^\ell \subset \J$
is defined as the vector space of test jets with the following properties:
\begin{itemize}[leftmargin=2em]
\item[\rm{(i)}] For all~$y \in M$ and all~$x$ in an open neighborhood of~$M$,
the directional derivatives
\beq \label{derex}
\big( \nabla_{1, \v_1} + \nabla_{2, \v_1} \big) \cdots \big( \nabla_{1, \v_p} + \nabla_{2, \v_p} \big) \L(x,y)
\eeq
(computed componentwise in charts around~$x$ and~$y$)
exist for all~$p \in \{1, \ldots, \ell\}$ and all~$\v_1, \ldots, \v_p \in \J^\ell$.
\item[\rm{(ii)}] The functions in~\eqref{derex} are $\rho$-integrable
in the variable~$y$, giving rise to locally bounded functions in~$x$. More precisely,
these functions are in the space
\[ L^\infty_\text{\rm{loc}}\Big( L^1\big(M, d\rho(y) \big), d\rho(x) \Big) \:. \]
\item[\rm{(iii)}] Integrating the expression~\eqref{derex} in~$y$ over~$M$
with respect to the measure~$\rho$,
the resulting function (defined for all~$x$ in an open neighborhood of~$M$)
is continuously differentiable in the direction of every jet~$\u \in \Jtest$.
\end{itemize}
\end{Def} \noindent
Here and throughout this paper, we use the following conventions for partial derivatives and jet derivatives:
\begin{itemize}[leftmargin=2em]
\itemD Partial and jet derivatives with an index $i \in \{ 1,2 \}$, as for example in~\eqref{derex}, only act on the respective variable of the function $\L$.
This implies, for example, that the derivatives commute,
\[ 
\nabla_{1,\v} \nabla_{1,\u} \L(x,y) = \nabla_{1,\u} \nabla_{1,\v} \L(x,y) \:. \]
\itemD The partial or jet derivatives which do not carry an index act as partial derivatives
on the corresponding argument of the Lagrangian. This implies, for example, that
\[ \nabla_\u \int_\F \nabla_{1,\v} \, \L(x,y) \: d\rho(y) =  \int_\F \nabla_{1,\u} \nabla_{1,\v}\, \L(x,y) \: d\rho(y) \:. \]
\end{itemize}
We point out that, in contrast to the method and conventions used in~\cite{jet},
{\em{jets are never differentiated}}.

The combination of derivatives in~\eqref{derex} requires a brief explanation.
In the case~$p=1$, the combination of directional derivatives in~\eqref{derex} is defined by
\[ \big( D_{1, v} + D_{2, v} \big) \L(x,y) := \frac{d}{d\tau} 
\L\big( F_\tau(x), F_\tau(y) \big) \big|_{\tau=0} \:,\]
where~$F_\tau$ is the flow of the vector field~$v$.
The higher derivatives are defined inductively. However, we use the convention
that the partial derivatives act only on the arguments of~$\L$, but not on any other jets.
This means that one must subtract the terms involving derivatives of the jets. For example,
\[ \big( D_{1, v} + D_{2, v} \big)^2 \L(x,y) := \frac{d^2}{d\tau^2} 
\L\big( F_\tau(x), F_\tau(y) \big) \big|_{\tau=0} 
- \big( D_{1, D_v v} + D_{2, D_v v} \big) \L(x,y) \:,\]
and similarly for higher derivatives.
The condition in Definition~\ref{defJvary}~(i) implies that all the resulting terms
must exist.

Linearized solutions are linear perturbations of~$\rho$ which
preserve the weak EL equations~\eqref{ELtest}.
We now give the precise definition (for more details see~\cite[Section~4.2]{jet}).
\begin{Def} \label{deflinear}
A jet~$\v \in \J^1$ is referred to as a {\bf{solution of the linearized field equations}} if
\[ \nabla_{\u} \int_M \big( \nabla_{1,\v} + \nabla_{2,\v}\big) \L(x,y) \: d\rho(y) = 
\nabla_{\u} \nabla_{\v}\:\s \qquad \text{for all~$\u \in \Jtest$ and~$x \in M$}\:. \]
The vector space of all linearized solutions is denoted by~$\Jlin \subset \J^1$.
\end{Def} \noindent

\section{The Abstract Perturbation Expansion} \label{secabstract}

\subsection{Perturbation Expansion for the Universal Measure} \label{secrhopert}
Let~$\rho$ be a measure (not necessarily a critical point of the causal variational principle).
We want to construct a measure~$\tilde{\rho}$ which satisfies the weak EL equations.
To this end, we make the ansatz
\beq \label{rhotilde}
\tilde{\rho} = F_* \big( f \, \rho \big) \:,
\eeq
where~$f$ and~$F$ are smooth,
\beq \label{fFdef}
f \in C^\infty\big(M,\R^+ \big) \qquad \text{and} \qquad
F \in C^\infty\big(M, \F \big)
\eeq
(where smooth on~$M$ again means that there exists a smooth extension to~$\F$).
This ansatz is motivated mainly by its simplicity.
More general perturbations of the universal measure will be studied in Section~\ref{secfragment}.

We denote the test space for the measure $\tilde\rho$ by $\tilde{\J}^\text{\tiny{test}}$, i.e.\
\[ \tilde{\J}^\text{\tiny{test}} \subset \big\{ \u = (a,u) \text{ with } a \in C^\infty(\tilde{M}, \R) \text{ and }
u \in C^\infty(\tilde{M}, T\F) \big\} \:, \]
where~$\tilde{M}:= \supp \tilde{\rho}$ is the support of the varied measure.
We write the weak EL equations~\eqref{ELtest} for the measure~$\tilde\rho$ as
\beq \label{ELtest0}
\nabla_{1,\tilde{\u}(F(x))}  \bigg( \int_M \L\big(F(x), F(y) \big)\: f(y) \, d\rho(y) - \s \bigg) = 0 \qquad \text{for all~$\tilde{\u} \in \tilde{\J}^\text{\rm{\tiny{test}}}$} \:,
\eeq
to be evaluated pointwise for all~$x \in M$. Here the notation $\nabla_{1,\tilde{\u}}$
clarifies that the derivative acts on the first argument of the Lagrangian.
On the constant~$\s$ it acts by multiplication with the scalar component,
\[ \nabla_{1,\tilde{\u}(F(x))} \,\s = \nabla_{\tilde{\u}(F(x))} \,\s
= a\big(F(x)\big)\, \s \:, \]
where we again denote the components by~$\u=(a,u)$.
Note that, being defined on~$\tilde{M}$, the jet~$\tilde{\u}$ can be evaluated at~$x \in M$ 
only after composing it with~$F$.
In order to rewrite this equation in a way where~$x$ and~$y$ are treated in a more symmetric way,
we multiply~\eqref{ELtest0} by the function~$f(x)$ and write this function inside the
brackets,
\[ 
\nabla_{1,\tilde{\u}(F(x))} \bigg( \int_M f(x) \:\L\big(F(x), F(y) \big)\: f(y) \, d\rho(y) - 
\s\:f(x) \bigg) = 0 \qquad \text{for all~$\tilde{\u} \in \tilde{\J}^\text{\rm{\tiny{test}}}$} \:. \]
Working in charts makes it possible to identify the tangent spaces at different
points simply by identifying the components. In particular, we use this
method in order to identify~$\tilde{\u}(F(x))$ with a jet~$\u(x)$.
We choose the jet space~$\tilde{\J}^\text{\rm{\tiny{test}}}$ such that,
under this identification, it coincides with~$\Jtest$.
Then the weak EL equations can be written as
\beq \label{final}
\nabla_{1,\u} \bigg( \int_M f(x) \:\L\big(F(x), F(y) \big)\: f(y)\: d\rho(y)
-\s\: f(x) \bigg) = 0 \:,
\eeq
to be satisfied for all~$\u \in \Jtest$ and all~$x \in M$.
We again point out that the derivative~$\nabla_{1,\u}$ acts on the first argument of the Lagrangian
and on the constant~$\s$, but the factor~$f(x)$ is not differentiated.

In physical applications, it is relatively easy to construct an approximate solution of the
EL equations (typically by regularizing Dirac sea structures in the presence of a classical bosonic
potential; for details see~\cite{cfs}). With this in mind, we
now assume that the measure~$\rho$ is close to a critical point in the sense that
\beq \label{E1d}
\nabla_{\u} \bigg( \int_{M} \L(x,y)\: d\rho(y) -\s \bigg) = \lambda \,\nabla_{\u} E^{(1)}
\eeq
with an error term $E^{(1)}$, where~$\lambda \in \R$ is a small parameter.
We expand both~$f$ and~$F$ in a power series in~$\lambda$.
For the function~$f$, we make the perturbation ansatz
\beq \label{fxser}
f(x) = \sum_{p=0}^\infty \lambda^p\: f^{(p)}(x)  \qquad \text{with} \qquad f^{(0)}(x) = 1 \:,
\eeq
where the choice of~$f^{(0)}$ will ensure that the measure~$\tilde{\rho}$ goes over to the unperturbed measure~$\rho$
in the limit~$\lambda \rightarrow 0$. For the expansion of~$F$, we choose a chart around~$x$ and
write~$F(x)$ in components as~$(F(x)^\alpha)_{\alpha=1,\ldots, m}$. Then we can
expand~$F$ componentwise,
\beq \label{Fxser}
F(x)^\alpha = \sum_{p=0}^\infty \lambda^p\: F^{(p)}(x)^\alpha
\qquad \text{with} \qquad F^{(0)}(x)^\alpha = x^\alpha \:.
\eeq
For ease in notation, we shall omit the index~$\alpha$ from now on.
But one should keep in mind that the expansion of~$F(x)$ always involves the choice of a chart
around~$x$.

In the next lemma we evaluate~\eqref{final} to any order~$p=1,2,\ldots$ in~$\lambda$.
In order to simplify the combinatorics, it turns out to be convenient to work instead of
the function~$f$ with its logarithm
\beq \label{bdef}
c(x) := \log f(x) \:,
\eeq
which, similar to~\eqref{fxser}, we again expand in powers of~$\lambda$,
\beq \label{bser}
c(x) = \sum_{p=0}^\infty \lambda^p\: c^{(p)}(x) \qquad \text{with} \qquad c^{(0)}(x) = 0 \:.
\eeq
Moreover, we combine the~$c^{(p)}$ and~$F^{(p)}$ to jets~$\w^{(p)}$, i.e.
\beq \label{vpdef}
\w^{(p)} := \big( c^{(p)}, F^{(p)} \big) \qquad \text{for~$p=1,2,\ldots$}\:.
\eeq

\begin{Lemma} \label{lemmapert}
To every order $p=1,2,\ldots$, the weak EL equations~\eqref{final} can be written as
\beq \label{perteq}
\begin{split}
0&=\nabla_{\u}
\sum_{\ell=1}^p \frac{1}{\ell!} \sum_{\stackrel{\text{\scriptsize{
$q_1, \ldots, q_\ell \geq 1$}}}{\text{with } q_1+\cdots+q_\ell=p}}\\
& \qquad \times \bigg\{ \int_{M} \big(\nabla_{1,\w^{(q_1)}} + \nabla_{2, \w^{(q_1)}} \big) \cdots 
\big(\nabla_{1, \w^{(q_\ell)}} + \nabla_{2, \w^{(q_\ell)}} \big) \L(x,y) \:d\rho(y) \\
&\qquad\qquad -\s\: c^{(q_1)}(x) \cdots c^{(q_\ell)}(x) \bigg\} \:.
\end{split}
\eeq
\end{Lemma}
\Proof The combinatorics can be handled elegantly by working with exponentials.
We explain the method in the example of a function~$h(F(x))$. We first expand in
a Taylor series,
\[ h \big(F(x) \big) = h \big(x + (F(x)-x) \big) 
= \sum_{p=0}^\infty \frac{1}{p!}\: D^p_{F(x)-x} h(x) 
= \exp \Big( D^p_{F(x)-x} \Big) h(x) \:. \]
The exponential on the right simply is a short notation for the formal power series.
Multiplying by~$f(x)$ and using~\eqref{bdef}, we can combine the exponentials
to obtain a jet derivative,
\[ f(x)\, h\big(F(x) \big) = e^{c(x)}\, \exp \Big( D^p_{F(x)-x} \Big) h(x) 
= \exp \Big( \nabla^p_{\tilde{\w}} \Big) h(x) \:, \]
where the jet~$\tilde{\w}$ has the components
\[ \tilde{\w}(x) = \big( c(x), F(x)-x \big) \:. \]
Applying the same method to the integrand in~\eqref{final} gives
\beq \label{ffexp0}
f(x) \:\L\big(F(x), F(y) \big)\: f(y) = \exp \Big( \nabla_{1, \tilde{\w}} + \nabla_{2, \tilde{\w}} \Big)
\L(x,y)\:,
\eeq
where we used that the derivatives all act on the arguments of~$\L(x,y)$, making it possible to
simplify the prefactors with the usual computation rules of the exponential.
Using the abbreviation
\[ \overline{\nabla}_{\w}  := \big( \nabla_{1,\w} + \nabla_{2,\w} \big) \:, \]
the identity~\eqref{ffexp0} can be written in the compact form
\beq \label{ffexp}
f(x) \:\L\big(F(x), F(y) \big)\: f(y) = e^{\overline{\nabla}_{\tilde{\w}}}\: \L(x,y)\:.
\eeq

It remains to expand the exponential in~\eqref{ffexp} in powers of~$\lambda$.
Inserting the perturbation expansion of~$\tilde{\w}$, we obtain
\[ e^{\overline{\nabla}_{\tilde{\w}}}
 = \exp \big( \lambda \,\overline{\nabla}_{\w^{(1)}} + 
\lambda^2 \,\overline{\nabla}_{\w^{(2)}} + \cdots \big)\:. \]
Let us compute the~$p^\text{th}$ $\lambda$-derivative of this exponential at~$\lambda=0$.
We consider the contribution involving the factors~$\overline{\nabla}_{\w^{q_1}}, \ldots,
\overline{\nabla}_{\w^{q_\ell}}$. 
Since each factor~$\overline{\nabla}_{\w^{q}}$ comes with
a factor~$\lambda^q$, we clearly get a contribution only if~$q_1+ \cdots+ q_\ell = p$.
We thus obtain
\[ \frac{d^p}{d\lambda^p} e^{\overline{\nabla}_{\tilde{\w}}} \bigg|_{\lambda=0}
= \sum_{\ell=1}^p \sum_{\stackrel{\text{\scriptsize{
$q_1, \ldots, q_\ell \geq 1$}}}{\text{with } q_1+\cdots+q_\ell=p}} \!\!\!\!\!c_{q_1, \ldots, c_\ell} \;
\overline{\nabla}_{\w^{(q_1)}} \cdots \overline{\nabla}_{\w^{(q_\ell)}} \:, \]
where~$c_{q_1, \ldots, c_\ell}$ are combinatorial factors which can be determined as follows.
Clearly, each $\lambda$-derivative annihilates one of the factors~$\lambda$ of the
monomial~$\lambda^{q_1} \cdots \lambda^{q_\ell}$. We must count the number of possibilities
with which this can occur. We first distinguish those $\lambda$-derivatives which act on the exponential
according to
\[ \frac{d}{d\lambda}\: e^{\lambda^q\, \overline{\nabla}_{\w^{(q)}}} = q\, \lambda^{q-1}\, \overline{\nabla}_{\w^{(q)}}\:
e^{\lambda^q\, \overline{\nabla}_{\w^{(q)}}}\:. \]
Note that each such derivative generates a factor~$\overline{\nabla}_{\w^{(q)}}$.
We use the convention that, carrying out the $\lambda$-derivatives consecutively,
the first $\lambda$-derivative acting on the exponential generates the factor~$\overline{\nabla}_{\w^{(q_1)}}$,
the second such derivative generates the factor~$\overline{\nabla}_{\w^{(q_2)}}$, and so on.
Dropping this convention gives a factor~$1/\ell!$, i.e.
\[ \frac{d^p}{d\lambda^p} e^{\overline{\nabla}_{\tilde{\w}}} \bigg|_{\lambda=0}
= \sum_{\ell=1}^p \frac{1}{\ell!} \!\!\!\!\!\sum_{\stackrel{\text{\scriptsize{
$q_1, \ldots, q_\ell \geq 1$}}}{\text{with } q_1+\cdots+q_\ell=p}} \!\!\!\!\! \tilde{c}_{q_1, \ldots, c_\ell} \;
\overline{\nabla}_{\w^{(q_1)}} \cdots \overline{\nabla}_{\w^{(q_\ell)}} \]
with new combinatorial factors~$\tilde{c}_{q_1, \ldots, c_\ell}$ which are obtained 
simply by counting the number of possibilities of forming groups of $\lambda$-derivatives acting on~$\lambda^{q_1}$,
$\lambda^{q_2}$, and so on.
These combinatorial factors are given by the monomial theorem. We thus obtain
\begin{align*}
\frac{d^p}{d\lambda^p} e^{\overline{\nabla}_{\tilde{\w}}} \bigg|_{\lambda=0}
&= \sum_{\ell=1}^p \frac{1}{\ell!} \!\!\!\!\!\sum_{\stackrel{\text{\scriptsize{
$q_1, \ldots, q_\ell \geq 1$}}}{\text{with } q_1+\cdots+q_\ell=p}} \!\!\!\!\!
\begin{pmatrix} p \\ q_1 \cdots q_\ell \end{pmatrix} 
\bigg( \frac{d^{q_1}}{d\lambda^{q_1}} e^{\lambda^{q_1} \,\overline{\nabla}_{\w^{(q_1)}}} \bigg) 
\cdots \bigg( \frac{d^{q_\ell}}{d\lambda^{q_\ell}} e^{\lambda^{q_\ell} \,\overline{\nabla}_{\w^{(q_\ell)}}} \bigg)
\bigg|_{\lambda=0} \\
&= p! \:\sum_{\ell=1}^p  \frac{1}{\ell!} \sum_{\stackrel{\text{\scriptsize{
$q_1, \ldots, q_\ell \geq 1$}}}{\text{with } q_1+\cdots+q_\ell=p}}
\overline{\nabla}_{\w^{(q_1)}} \cdots \overline{\nabla}_{\w^{(q_\ell)}} \:.
\end{align*}
Using this formula in~\eqref{ffexp} gives
\[ \frac{1}{p!} \frac{d^p}{d\lambda^p} \Big( f(x) \:\L\big(F(x), F(y) \big)\: f(y) \Big)
=\sum_{\ell=0}^p \frac{1}{\ell !} \!\!\!\!\!\sum_{\stackrel{\text{\scriptsize{$q_1, \ldots, q_\ell \geq 1$}}} {\text{with } q_1+\cdots+q_\ell=p}} \!\!\!\!\! \overline{\nabla}_{\w^{(q_1)}} \cdots \overline{\nabla}_{\w^{(q_\ell)}} \, \L(x,y) \:. \]
Similarly, one derives the identity
\[ \frac{1}{p!} \frac{d^p}{d\lambda^p} \:e^{c(x)}\:\\
=\sum_{\ell=0}^p \frac{1}{\ell !} \!\!\!\!\!\sum_{\stackrel{\text{\scriptsize{$q_1, \ldots, q_\ell \geq 1$}}} {\text{with } q_1+\cdots+q_\ell=p}} \!\!\!\!\! c^{(q_1)}(x) \cdots c^{(q_\ell)}(x) \:. \]
Employing these formulas in~\eqref{final} gives the result.
\QED

\subsection{Green's Operators} \label{secgreen}
In Lemma~\ref{lemmapert} we rewrote the weak EL equations as the system of
equations~\eqref{perteq}, to be satisfied for every~$p=1,2,\ldots$. In order 
to solve this system of equations, we bring the contribution
involving~$\w^{(p)}$ to the left. We thus obtain the equation
\beq \label{deluM}
\nabla_{\u} \bigg( \int_{M} \big(\nabla_{1, \w^{(p)}} + \nabla_{2, \w^{(p)}} \big) \L(x,y) \:d\rho(y) 
-\s\: c^{(p)}(x) \bigg) = - \nabla_{\u} E^{(p)}(x) \:,
\eeq
where~$E^{(1)}$ is given by~\eqref{E1d}, whereas for~$p>1$ we have
\beq
\begin{split}
E^{(p)} &= \sum_{\ell=2}^p \frac{1}{\ell!} \sum_{\stackrel{\text{\scriptsize{
$q_1, \ldots, q_\ell \geq 1$}}}{\text{with } q_1+\cdots+q_\ell=p}} 
\bigg\{ -\s\: c^{(q_1)}(x) \cdots c^{(q_\ell)}(x) \\
& \qquad +\int_{M} \big(\nabla_{1,\w^{(q_1)}} + \nabla_{2, \w^{(q_1)}} \big) \cdots 
\big(\nabla_{1, \w^{(q_\ell)}} + \nabla_{2, \w^{(q_\ell)}} \big) \L(x,y) \:d\rho(y) \bigg\} \:. \label{Epdef}
\end{split}
\eeq
Before solving for~$\w^{(p)}$, we need to specify the jet space 
used for varying the measure: We denote the
continuous global one-jets of the cotangent bundle restricted to~$M$ by
\[ \J^* := C^0(M, \R) \oplus C^0(M, T^*\F) \:. \]
We let~$(\Jtest)^*$ be the quotient space
\begin{align*}
(\Jtest)^* &:= \J^* \Big/  \big\{ (g,\varphi) \in \J^* \:\big|\: g(x) \,a(x) + \la \varphi(x), u(x) \ra = 0 \\
&\qquad\qquad\qquad\qquad\quad  \text{ for all~$\u=(a,u) \in \Jtest$ and~$x \in M$} \big\} \:,
\end{align*}
where~$\la .,. \ra$ denotes the dual pairing of~$T^*_x\F$ and~$T_x\F$
(the reason for taking equivalence classes simply is that it is
convenient to disregard dual jets which are trivial on~$\Jtest$).
We thus obtain a mapping
\begin{align}
&\Delta_\ell \::\: \underbrace{\J^\infty \times \cdots \times \J^\infty}_{\text{$\ell$ factors}}
\rightarrow (\Jtest)^* \:, \label{Lapldef} \\
&\big\la \u, \Delta_\ell \big[ \v_1, \ldots, \v_\ell \big] \big\ra(x) = \frac{1}{\ell!} \:\nabla_{\u} \bigg( \int_{M} \big( \nabla_{1, \v_1} + \nabla_{2, \v_1} \big) \cdots \big( \nabla_{1, \v_\ell} + \nabla_{2, \v_\ell} \big) \L(x,y)\: d\rho(y)\notag \\
&\qquad\qquad\qquad\qquad\qquad\quad\qquad\qquad  -\s\: b_1(x) \cdots b_\ell(x) \bigg) \:, \notag
\end{align}
valid for any~$\u \in \Jtest$.
We remark for clarity that the mapping~$\Delta_\ell$ is symmetric in its~$\ell$ arguments.
Choosing~$\ell=1$, we obtain the mapping~$\Delta \equiv \Delta_1 : \J^\infty \rightarrow (\Jtest)^*$ given by
\[ 
\la \u, \Delta \v \ra(x) = \nabla_\u \bigg( \int_{M} \big( \nabla_{1, \v} + \nabla_{2, \v} \big) \L\big(x, y \big)
\: d\rho(y) - \nabla_\v \:\s \bigg) \:. \]

\begin{Def} \label{defS}
A linear mapping~$S : (\Jtest)^* \rightarrow \J^\infty$ is referred to as a {\bf{Green's operator}} if
\beq \label{Sdefine}
\Delta\,S \, \v = -\v \qquad \text{for all~$\v \in (\Jtest)^*$} \:.
\eeq
\end{Def} \noindent
Clearly, a Green's operator exists if and only if the mapping~$\Delta$ is surjective.
In analogy to the situation for hyperbolic PDEs, the Green's operators need not be unique.
Indeed, just as in classical field theory, the difference of two Green's operators
is a solution of the linearized field equations (see Definition~\ref{deflinear}).
We remark that, similar as in classical field theory and quantum field theory,
one could work with specific Green's operators 
determined by support properties (like retarded or advanced Green's operators) or
by microlocal properties (like the Feynman propagator).
However, at this stage, where we merely seek for solutions of the weak EL equations
without specifying initial conditions, we cannot and need not specify the Green's operators.

With the above notions, we can write~\eqref{deluM} as
\beq \label{wpcond}
\Delta \w^{(p)} = -E^{(p)} \in (\Jtest)^* \:.
\eeq
Having a Green's operator to our disposal, we can solve this equation for~$\w^{(p)}$,
\beq \label{vpertprelim}
\w^{(p)} = S \, E^{(p)} \:.
\eeq
Combining this equation with~\eqref{E1d} and~\eqref{Epdef}, we have obtained
an iterative procedure for constructing measures which satisfy the weak EL equations~\eqref{final}.
We again point out that the Green's operator~$S$ is not unique. Indeed, there is the freedom
to choose a different Green's operator to every order in perturbation theory.
Exactly as in the analogous situation for hyperbolic PDEs, taking this freedom into account
gives rise to the {\em{general}} solution to the weak EL equations.
In order to make this non-uniqueness manifest, we prefer to write~\eqref{vpertprelim} as
\beq \label{vpert}
\w^{(p)} = S^{(p)} \, E^{(p)} \:,
\eeq
where~$S^{(1)}, S^{(2)}, \ldots$ are arbitrary Green's operators.

\subsection{Diagrams and Feynman Rules} \label{secdiagram}
We now summarize the above construction and formulate it in a
diagrammatic language. For simplicity, we leave out the parameter~$\lambda$,
which was used merely as a book-keeping device
in order to keep track of the different orders in perturbation theory.
We introduce the operators $\Delta_\ell$ by (see~\eqref{Lapldef})
\begin{align}
\Delta_0(x) &= \int_{M} \L(x,y)\: d\rho(y) -\s \label{delta0} \\
\Delta_\ell \big[\w_1, \ldots, \w_\ell \big](x)
&= \frac{1}{\ell!} \bigg( \int_{M} \big( \nabla_{1, \w_1} + \nabla_{2, \w_1} \big) \cdots \big( \nabla_{1, \w_\ell} + \nabla_{2, \w_\ell} \big) \L(x,y)\: d\rho(y) \notag \\
&\qquad\qquad\quad -\s\: c_1(x) \cdots c_\ell(x) \bigg) \qquad \text{(for $\ell \geq 1$)} \label{deltal}
\end{align}
and choose Green's operators~$S^{(p)}$ with~$p=1,2,\ldots$ as minus the inverse of~$\Delta \equiv \Delta_1$
(see Definition~\ref{defS}),
\beq \label{Sdef}
\Delta \,S^{(p)} \,\v = -\v \qquad \text{for all~$v \in (\Jtest)^*$}\:.
\eeq
Then the jets~$\w^{(p)}$ are defined iteratively by (see~\eqref{vpert})
\beq \label{vpiter}
\w^{(p)} = S^{(p)} \,E^{(p)} \:,
\eeq
where~$E^{(p)}$ depends on the previous jets~$\w^{(1)}, \ldots, \w^{(p-1)}$ by
(see~\eqref{E1d} and~\eqref{Epdef})
\begin{align}
E^{(1)}(x) &= \Delta_0(x) \label{E1def} \\
E^{(p)}(x) &= \sum_{\ell=2}^p E_\ell^{(p)}(x) \qquad \text{(for $p \geq 2$)} \label{Ep} \\
E_\ell^{(p)}(x) &= \sum_{\stackrel{\text{\scriptsize{
$q_1, \ldots, q_\ell \geq 1$}}}{\text{with } q_1+\cdots+q_\ell=p}}\;\;
\Delta_\ell \big[ \w^{(q_1)}, \ldots, \w^{(q_\ell)} \big](x) \:. \label{Elp}
\end{align}

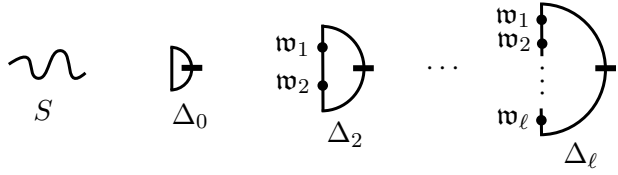
\begin{figure}
\begin{center}
{
\begin{pspicture}(-1,-0.5986111)(10.481037,1.0986111)
\psline[linecolor=black, linewidth=0.04](4.1838145,0.8036111)(4.1888146,-0.3563889)
\rput[bl](2.181037,-0.57361114){\normalsize{$\Delta_0$}}
\rput[bl](5.556037,0.13138889){\normalsize{$\cdots$}}
\psdots[linecolor=black, dotsize=0.14](7.096037,0.86972225)
\psbezier[linecolor=black, linewidth=0.04](1.0313146,0.16111112)(0.73074186,-0.101102486)(0.8845578,0.42225006)(0.7113146,0.456111111111112)(0.5380715,0.48997217)(0.55631465,0.056111112)(0.39131463,0.09111111)(0.22631463,0.1261111)(0.42631465,0.5611111)(0.01131464,0.27611113)
\psarc[linecolor=black, linewidth=0.04, dimen=outer](2.1588147,0.2186111){0.285}{-90.0}{90.0}
\psarc[linecolor=black, linewidth=0.04, dimen=outer](4.1713147,0.2186111){0.5625}{-90.0}{90.0}
\psdots[linecolor=black, dotsize=0.14](4.181037,0.4997222)
\psdots[linecolor=black, dotsize=0.14](4.186037,-0.0052777776)
\rput[bl](4.241037,-0.8186111){\normalsize{$\Delta_2$}}
\psline[linecolor=black, linewidth=0.08](4.5888147,0.22361112)(4.863815,0.22361112)
\rput[bl](3.551037,0.3913889){\normalsize{$\w_1$}}
\rput[bl](3.561037,-0.108611114){\normalsize{$\w_2$}}
\rput[bl](6.4860367,0.7713889){\normalsize{$\w_1$}}
\rput[bl](0.34103686,-0.4586111){\normalsize{$S$}}
\psline[linecolor=black, linewidth=0.04](2.1788146,0.5036111)(2.1788146,-0.06638889)
\psline[linecolor=black, linewidth=0.08](2.3088145,0.22861111)(2.5838146,0.22861111)
\psarc[linecolor=black, linewidth=0.04, dimen=outer](7.073815,0.20861112){0.87}{-90.0}{90.0}
\psline[linecolor=black, linewidth=0.04](7.093815,1.0786111)(7.0988145,0.3836111)
\rput{-90.0}(6.729648,7.2924256){\rput[bl](7.011037,0.28138888){\normalsize{$\cdots$}}}
\psline[linecolor=black, linewidth=0.04](7.0888147,-0.31138888)(7.093815,-0.6563889)
\psdots[linecolor=black, dotsize=0.14](7.101037,0.5497222)
\psdots[linecolor=black, dotsize=0.14](7.091037,-0.4702778)
\rput[bl](6.5060367,0.4463889){\normalsize{$\w_2$}}
\rput[bl](6.501037,-0.57861114){\normalsize{$\w_\ell$}}
\rput[bl](7.386037,-1.0986111){\normalsize{$\Delta_\ell$}}
\psline[linecolor=black, linewidth=0.08](7.8088145,0.22361112)(8.083815,0.22361112)
\end{pspicture}
}
\end{center}
\caption{Building blocks of Feynman diagrams.}
\label{figfeyn1}
\end{figure}

The universal measure~$\tilde{\rho}$ is obtained 
by (see~\eqref{rhotilde}, \eqref{fxser}, \eqref{Fxser}, \eqref{bdef} and~\eqref{vpdef})
\beq \label{rhoser}
\tilde{\rho} = F_* \big( e^c \, \rho \big) \qquad \text{where} \qquad
(c, F)(x) = (0,x) + \sum_{p=1}^\infty \w^{(p)}(x) \:.
\eeq
For the graphical representation, we denote the Green's operator by a wiggled line
and the operators~$\Delta_\ell$ by semicircles (see Figure~\ref{figfeyn1}).
Then the contributions to the perturbation expansion can be depicted by
Feynman diagrams as illustrated in Figure~\ref{figfeyn2}.
The combinatorics is given in~\eqref{Ep}, \eqref{Elp} and~\eqref{rhoser}.
We point out that our perturbation expansion only involves tree diagrams.

\begin{Remark} {\bf{(alternative form of the perturbation expansion)}} \label{rempertalt}
{\em{ For completeness, we now give an alternative form of the perturbation expansion 
will be used in Section~\ref{secfragment} and might
be useful for future applications. Namely, dividing by~$f(x)$, the weak EL equations~\eqref{final} can be written
alternatively as
\beq \label{finalbreve}
\nabla_{1,\u} \bigg( \int_M \L\big(F(x), F(y) \big)\: f(y)\: d\rho(y) -\s \bigg) = 0 \:,
\eeq
to be satisfied for all~$\u \in \Jtest$.
Expanding the equations in this form, one obtains the same perturbation expansion
as above, except that the operator~$\Delta_\ell$ in~\eqref{deltal} is to be modified to
\beq
\breve{\Delta}_\ell \big[\w_1, \ldots, \w_\ell \big](x) = \frac{1}{\ell!} \int_{M} \big( D_{1, w_1} + \nabla_{2, \w_1} \big) \cdots \big( D_{1, w_\ell} + \nabla_{2, \w_\ell} \big) \L(x,y)\: d\rho(y) \:. \label{deltabreve}
\eeq
This formulation has the advantage that the Lagrange multiplier~$\s$ drops out.
Moreover, it becomes clearer that the scalar component of the jets only enters at the point~$y$
(as is obvious in~\eqref{finalbreve} where only~$f(y)$ appears).
The disadvantage is that~\eqref{deltabreve} is less symmetric in the variables~$x$ and~$y$
(in particular, the form~\eqref{deltal} is of advantage for the derivation of conservation laws
for surface layer integrals in~\cite{jet,osi}).
}} \QEDrem
\end{Remark}
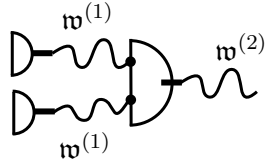
\begin{figure}
\begin{center}
{
\begin{pspicture}(-4,-0.5294921)(6.7622223,1.0294921)
\psline[linecolor=black, linewidth=0.04](1.595,0.6336675)(1.6,-0.52633244)
\psbezier[linecolor=black, linewidth=0.04](1.5725,0.32616752)(1.2719272,0.063953936)(1.4257431,0.5873065)(1.2525,0.621167534722224)(1.0792569,0.6550286)(1.0975,0.22116753)(0.9325,0.25616753)(0.7675,0.29116753)(0.9675,0.72616756)(0.5525,0.44116753)
\psarc[linecolor=black, linewidth=0.04, dimen=outer](0.0,0.43366754){0.285}{-90.0}{90.0}
\psarc[linecolor=black, linewidth=0.04, dimen=outer](1.5825,0.048667535){0.5625}{-90.0}{90.0}
\psdots[linecolor=black, dotsize=0.14](1.5922222,0.32977864)
\psdots[linecolor=black, dotsize=0.14](1.5972222,-0.17522135)
\psline[linecolor=black, linewidth=0.08](2.0,0.053667534)(2.275,0.053667534)
\rput[bl](0.67722225,0.77144533){\normalsize{$\w^{(1)}$}}
\psline[linecolor=black, linewidth=0.04](0.04,-0.07133246)(0.04,-0.64133245)
\psline[linecolor=black, linewidth=0.08](0.28,0.44366753)(0.555,0.44366753)
\psbezier[linecolor=black, linewidth=0.04](1.5775,-0.16487414)(1.3319272,-0.5762938)(1.4007431,-0.20915505)(1.2225,-0.17399320233745413)(1.0442568,-0.13883136)(1.1375,-0.47618383)(0.9575,-0.44853833)(0.7775,-0.42089286)(0.8975,-0.11052197)(0.5625,-0.34747434)
\rput[bl](0.6422222,-0.9535547){\normalsize{$\w^{(1)}$}}
\psline[linecolor=black, linewidth=0.08](0.31,-0.33133247)(0.585,-0.33133247)
\psarc[linecolor=black, linewidth=0.04, dimen=outer](0.025,-0.34633246){0.285}{-90.0}{90.0}
\psline[linecolor=black, linewidth=0.04](0.02,0.72866756)(0.02,0.15866753)
\psbezier[linecolor=black, linewidth=0.04](3.2725,-0.068832465)(2.9719272,-0.33104607)(3.1257432,0.19230647)(2.9525,0.22616753472222398)(2.7792568,0.2600286)(2.7975,-0.17383246)(2.6325,-0.13883246)(2.4675,-0.10383247)(2.6675,0.33116755)(2.2525,0.046167534)
\rput[bl](2.7222223,0.3564453){\normalsize{$\w^{(2)}$}}
\end{pspicture}
}
\end{center}
\caption{A simple Feynman diagram.}
\label{figfeyn2}
\end{figure}

\subsection{Constructing Nonlinear Solutions of the Field Equations} \label{seclinpert}
We now explain how the general construction of Section~\ref{secrhopert} can be
adapted in order to construct nonlinear solutions of the field equations.
We consider the setting that~$\rho$ is a minimizing measure, and we again assume that we
are given a Green's operator~$S$ (see Definition~\ref{defS}).
Moreover, we are given a jet~$\w^{(1)} \in \J^\infty$ being
a solution of the linearized field equations (see Definition~\ref{deflinear}).
Our goal is to construct a family of solutions~$(\tilde{\rho}_\tau)_{\tau \in \R}$ of the
weak EL equations of the form~\eqref{rhotilde} whose first variation coincides with~$\w^{(1)}$, i.e.
\[ \tilde{\rho}_\tau|_{\tau=0} = \rho \qquad \text{and} \qquad
\big( \partial_\tau f, \partial_\tau F \big) \big|_{\tau=0} = \w^{(1)}\:. \]
To this end, we construct the jets~$\w^{(2)}, \w^{(3)}, \ldots$ iteratively
again by~\eqref{vpiter}, \eqref{Ep} and~\eqref{Elp} with~$\Delta_\ell$ according to~\eqref{deltal}
(note that now~$\Delta_0=\Delta_1[\w^{(1)}]=0$).
The desired family of measures~$(\tilde{\rho}_\tau)$ is then defined 
similar to~\eqref{rhoser} by inserting powers of~$\tau$, i.e.
\[ \tilde{\rho}_\tau = (F_\tau)_* \big( e^{c_\tau} \, \rho \big) \qquad \text{with} \qquad
(c_\tau, F_\tau)(x) = (0,x) + \sum_{p=1}^\infty \tau^p\, \w^{(p)}(x)
\:. \]

\subsection{Perturbing a Vacuum Measure} \label{secvacpert}
In the applications, one often knows a critical measure
which typically describes the vacuum of the system. Then the system is modified,
for example by introducing particles and/or fields. The task is to construct
a solution of the weak EL equations, starting from the modified system.
We now adapt the construction of Section~\ref{secrhopert}
to this setting. To this end, we assume that~$\rho$ is a measure
which satisfies the EL equations~\eqref{EL1}.
Moreover, we assume that we are given a Green's operator~$S$ (see Definition~\ref{defS}).
The modified system is described by a measure~$\hat{\rho}$ which, in analogy
to~\eqref{rhotilde}, we assume to be of the form
\beq \label{rhohat}
\hat{\rho} = H_* \big( h \, \rho \big) \:,
\eeq
where~$h$ and~$H$ are smooth,
\[ h \in C^\infty\big(M,\R^+ \big) \qquad \text{and} \qquad
H \in C^\infty\big(M, \F \big) \:. \]
Clearly, the measure~$\hat{\rho}$ is no longer a solution of the weak EL equations.
Similar to~\eqref{bdef}, \eqref{vpdef} and~\eqref{rhoser},
we expand~$h$ and~$H$ and rewrite the coefficients with jets,
\beq \label{wpex}
\big( \log h, H \big)(x)  =  (0,x) + \sum_{p=1}^\infty \v^{(p)}(x) \qquad \text{with} \qquad
\v^{(p)} \in \J^\infty\:.
\eeq
Here we need to assume that the resulting jets~$\v^{(p)}$ are in~$\J^\infty$.

In order to construct a corresponding solution of the EL equations,
we again make the ansatz~\eqref{rhotilde} and describe~$f$ and~$F$
by jets~$\w^{(p)}$ (see~\eqref{Fxser}, \eqref{bdef}, \eqref{bser}
and~\eqref{vpdef}).
Now we perform the perturbation expansion similar to~\eqref{Ep}--\eqref{rhoser},
taking into account the inhomogeneity~$\v^{(p)}$ to every order in perturbation theory.
More precisely, \eqref{vpiter} is to be replaced by
\[ 
\w^{(p)} = \v^{(p)} + S^{(p)} \,\Big( E^{(p)} + \Delta \v^{(p)} \Big) \:. \]
Indeed, applying the operator~$\Delta$ and using the defining equation of the 
Green's operator~\eqref{Sdefine}, one sees that the
relations~\eqref{wpcond} again hold.

\section{Perturbation Theory with Fragmentation} \label{secfragment}
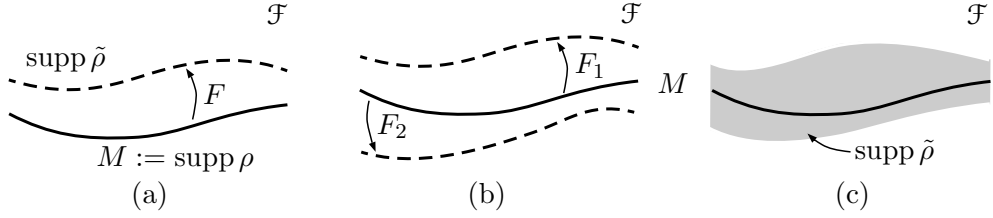
\begin{figure}
\begin{center}
{
\begin{pspicture}(-0.7,-1.3734679)(14.770368,1.3734679)
\definecolor{colour0}{rgb}{0.8,0.8,0.8}
\pspolygon[linecolor=colour0, linewidth=0.02, fillstyle=solid,fillcolor=colour0](9.339256,-0.2956901)(9.428145,-0.33124566)(9.592589,-0.38902342)(9.788145,-0.42902344)(10.001478,-0.45569012)(10.27259,-0.45124567)(10.543701,-0.44235677)(10.765923,-0.41124567)(11.108145,-0.35346788)(11.339256,-0.30013454)(11.548145,-0.24235678)(11.881478,-0.15791233)(12.094811,-0.10902344)(12.35259,-0.055690106)(12.659256,0.0020876736)(12.877034,0.033198785)(13.041478,0.046532117)(13.045923,0.6287543)(12.721478,0.695421)(12.374812,0.77097654)(12.0148115,0.8331988)(11.748145,0.8643099)(11.3437,0.8643099)(11.050367,0.8331988)(10.717034,0.7443099)(10.388145,0.6287543)(10.103701,0.5531988)(9.725923,0.49097657)(9.4725895,0.53097653)(9.3437,0.57986546)
\rput[bl](1.1703675,-0.8623568){$M:=\supp \rho$}
\psbezier[linecolor=black, linewidth=0.04, linestyle=dashed, dash=0.17638889cm 0.10583334cm](0.0059230137,0.3643099)(0.80589837,0.1160612)(1.3220041,0.2707834)(1.8170341,0.42653211805555347)(2.3120642,0.5822808)(3.01753,0.7522001)(3.705923,0.48430988)
\psbezier[linecolor=black, linewidth=0.04](0.009256347,-0.086801216)(0.5570095,-0.37171656)(0.99755967,-0.43366104)(1.6670341,-0.40124565972222515)(2.3365085,-0.36883026)(2.7330856,-0.07335544)(3.7092564,0.033198785)
\rput[bl](1.6325897,-1.3734679){(a)}
\rput[bl](6.1037006,-1.3690234){(b)}
\rput[bl](10.988145,-1.3690234){(c)}
\rput[bl](0.23703413,0.4665321){$\supp \tilde{\rho}$}
\psbezier[linecolor=black, linewidth=0.02, arrowsize=0.05291667cm 2.0,arrowlength=1.4,arrowinset=0.0]{->}(2.4488256,-0.16235676)(2.4691994,-0.06597727)(2.505363,0.112867646)(2.494812,0.19444763688016337)(2.4842608,0.27602762)(2.423824,0.42073712)(2.3614786,0.5176432)
\rput[bl](2.5881453,0.039865453){$F$}
\psbezier[linecolor=black, linewidth=0.04](4.6714787,0.22875434)(5.2192316,-0.056161016)(5.659782,-0.11810549)(6.3292565,-0.08569010416666856)(6.9987307,-0.053274717)(7.3953075,0.24220012)(8.371479,0.34875435)
\psbezier[linecolor=black, linewidth=0.04, linestyle=dashed, dash=0.17638889cm 0.10583334cm](4.677034,0.675421)(5.4770093,0.4271723)(5.9931154,0.5818945)(6.4881454,0.7376432291666666)(6.9831753,0.89339197)(7.688641,1.0633112)(8.377034,0.795421)
\psbezier[linecolor=black, linewidth=0.04, linestyle=dashed, dash=0.17638889cm 0.10583334cm](4.690367,-0.59569013)(5.48746,-0.84393877)(6.466691,-0.50699437)(6.9555087,-0.3423567708333326)(7.4443264,-0.17771916)(7.7311215,0.085533455)(8.319257,-0.06680121)
\rput[bl](4.899256,-0.41346788){$F_2$}
\rput[bl](7.5303674,0.395421){$F_1$}
\psbezier[linecolor=black, linewidth=0.02, arrowsize=0.05291667cm 2.0,arrowlength=1.4,arrowinset=0.0]{->}(7.3821588,0.18875434)(7.4025326,0.28513384)(7.4386964,0.46397877)(7.4281454,0.5455587479912742)(7.417594,0.62713873)(7.357157,0.77184826)(7.2948117,0.8687543)
\psbezier[linecolor=black, linewidth=0.02, arrowsize=0.05291667cm 2.0,arrowlength=1.4,arrowinset=0.0]{->}(4.7941318,0.086532116)(4.7737575,-0.009847382)(4.759816,-0.113136746)(4.7703676,-0.1947167341023888)(4.7809186,-0.27629673)(4.8413553,-0.4876729)(4.881479,-0.5934679)
\psbezier[linecolor=black, linewidth=0.04](9.355923,0.22430989)(9.903676,-0.06060546)(10.344226,-0.122549936)(11.0137005,-0.09013454861111199)(11.683175,-0.05771916)(12.079752,0.23775567)(13.055923,0.3443099)
\psbezier[linecolor=white, linewidth=0.04](9.315923,-0.27791232)(9.863676,-0.5628277)(10.570893,-0.46477216)(11.004812,-0.3879123263888903)(11.438731,-0.3110525)(12.070864,-0.060022105)(13.047034,0.046532117)
\psbezier[linecolor=white, linewidth=0.04](9.302589,0.6065321)(9.850343,0.32161677)(10.55756,0.7396723)(11.080367,0.8298654513888895)(11.603175,0.9200586)(12.168641,0.851089)(13.038145,0.6287543)
\rput[bl](11.250367,-0.7601346){$\supp \tilde{\rho}$}
\psbezier[linecolor=black, linewidth=0.02, arrowsize=0.05291667cm 2.0,arrowlength=1.4,arrowinset=0.0]{->}(11.195492,-0.5934679)(11.004142,-0.572549)(10.904003,-0.54737675)(10.819256,-0.4989638143801977)(10.734508,-0.45055088)(10.681424,-0.40389943)(10.622507,-0.34902343)
\rput[bl](8.628145,0.19542101){$M$}
\rput[bl](3.4370341,1.1065321){$\F$}
\rput[bl](8.143701,1.115421){$\F$}
\rput[bl](12.739256,1.1109766){$\F$}
\end{pspicture}
}
\end{center}
\caption{Fragmentation of the measure~$\rho$.}
\label{figfragment}
\end{figure}

The perturbation expansion of the previous section was based on the ansatz
that the perturbed measure~$\tilde{\rho}$ should be of the form~\eqref{rhotilde}
with~$f$ and~$F$ according to~\eqref{fFdef}. Intuitively speaking, this ansatz means
that the support of the measure is changed smoothly as a whole
(see Figure~\ref{figfragment}~(a)),
but it is impossible
to model a situation where the measure~$\rho$ ``disintegrates'' into several
``components'' which are perturbed differently (see Figure~\ref{figfragment}~(b)).
We now extend the constructions Section~\ref{secabstract} such as to allow
for such a so-called {\em{fragmentation}} of the universal measure.

We consider the following setting. Similar as in Section~\ref{seclinpert}
we want to construct nonlinear solutions of the field equations.
Therefore, we assume that~$\rho$ is a measure which satisfies the
weak EL equations~\eqref{ELtest}.
We choose a parameter~$L \in \N$ and consider mappings
\[ f_\as \in C^\infty\big(M,\R^+ \big)\:,\quad
F_\as \in C^\infty\big(M, \F \big) \qquad \text{with~$\as = 1, \ldots, L$}\:. \]
For the {\em{universal measure with fragmentation}} we make the ansatz
\beq \label{rhotildea}
\tilde{\rho} = \frac{1}{L} \sum_{\as=1}^L (F_\as)_* \big( f_\as \, \rho \big) \:.
\eeq
We refer to~$L$ as the {\em{number of subsystems}} and to~$\as$ as the {\em{subsystem index}}.
Clearly, for one subsystem, \eqref{rhotildea} reduces to our earlier ansatz~\eqref{rhotilde}.
The larger~$L$ is chosen, the more freedom we have in perturbing the measure~$\rho$.
We point out that we may choose~$L$ arbitrarily large. In the limit~$L \rightarrow \infty$, one can
even describe situations where the support of the measure~$\rho$ is ``enlarged'' by the perturbation
as shown in Figure~\ref{figfragment}~(c). We also note that a universal
measure of the form~\eqref{rhotildea} is closely related to the mechanism of microscopic mixing
as introduced in~\cite{qft}; this will be explained further in Section~\ref{secmixing}.

\subsection{Linearized Field Equations for Fluctuations} \label{seclinfluct}
It is a bit easier to perform the perturbation expansion with fragmentation
in the alternative formulation introduced in Remark~\ref{rempertalt}, because
then the scalar component of the jets appears only as a function of the variable~$y$
(but of course, all our results can be rewritten in a straightforward way
in the formulation~\eqref{final}).
Adapted to the measure~\eqref{rhotildea}, the weak EL equations~\eqref{finalbreve} read
\[ \nabla_{1,\u_\as} \bigg( \frac{1}{L} \sum_{\bs=1}^L \int_M \L\big(F_\as(x), F_\bs(y) \big)\: f_\bs(y)\: d\rho(y)
-\s \bigg) = 0 \:, \]
to be satisfied for all jets~$\u \in (\Jtest)^L$ as well as for all~$x \in M$ and~$\as \in \{1,\ldots, L\}$.
Since in finite dimension, pointwise evaluation is the same as weak evaluation, we can write
this equation equivalently as
\beq \label{prelimb}
\frac{1}{L} \sum_{\as=1}^L
\nabla_{1,\u_\as} \bigg( \frac{1}{L} \sum_{\bs=1}^L \int_M \L\big(F_\as(x), F_\bs(y) \big)\: f_\bs(y)\: d\rho(y)
-\s \bigg) = 0 \:,
\eeq
which must hold for all~$\u \in (\Jtest)^L$ and all~$x \in M$.

In preparation of the perturbation expansion, we derive the corresponding linearized
field equations. To this end, we again expand~$f$ and~$F$ according to~\eqref{fxser}
and~\eqref{Fxser}. To first order, the EL equations~\eqref{prelimb} become
\beq \label{error}
\begin{split}
&\big\la \u, \Delta \big[\w^{(1)} \big] \big\ra(x) := \frac{1}{L^2} \sum_{\as, \bs=1}^L \\
&\qquad \times \nabla_{\u_\as(x)} \int_M  \Big( \big(D_{1,w^{(1)}_\as} + D_{2,w^{(1)}_\bs} \big) \L(x, y) + \L(x,y)
\: f^{(1)}_\bs(y) \Big)\: d\rho(y)
\end{split}
\eeq
with~$\w^{(1)}_\as:= (f_\as^{(1)}, F^{(1)}_\as)$.
Note that the vector component of the jet~$\w^{(1)}_\as$ shifts the support of the universal measure in each
subsystem independently (as shown in Figure~\ref{figfragment}~(b)).

At this point, it is helpful to decompose
the jets into components independent of the subsystem index and
components whose mean vanishes, i.e.
\beq \label{decompose}
\u = \bar{\u} + \u_\Fluct \qquad \text{with} \qquad
\bar{\u}_\as(x) := \frac{1}{L} \sum_{\bs=1}^L \u_\bs(x) \:.
\eeq
Here the subscript ``F'' can be thought of as referring to the ``fragmentation''
of the universal measure or as describing the ``fluctuations'' of the jets in the
subsystems. For a convenient notation, we usually omit the subsystem index of~$\bar{\u}$.
The above splitting gives rise to a direct sum decomposition of
the jet spaces, which we write as
\[ \J^L = \bar{\J} \oplus \J_\Fluct \]
and similarly for the jet spaces~$\Jtest$ and~$\J^\infty$.

Using these notions, we can carry out the $\bs$-sum in~\eqref{error} to obtain
\[ \big\la \u, \Delta \big[\w^{(1)} \big] \big\ra(x) = 
\frac{1}{L} \sum_{\as=1}^L
\nabla_{\u_\as(x)} \int_M 
\Big( \big(D_{1,w^{(1)}_\as} + D_{2, \bar{w}^{(1)}} \big) \L(x, y)+ \L(x,y)\: \bar{f}^{(1)}(y) \Big)\: d\rho(y) \:. \]
The fluctuations drop out completely when testing in~$\bJtest$,
\[ 
\big\la \bar{\u}, \Delta \big[\w^{(1)} \big] \big\ra(x) = 
\nabla_{\bar{\u}(x)} \int_M 
\Big( \big(D_{1,\bar{w}^{(1)}} + D_{2, \bar{w}^{(1)}} \big) \L(x, y)+ \L(x,y)\: \bar{f}^{(1)}(y) \Big)\: d\rho(y) \:, \]
giving back the linearized field equations without fragmentation.
But clearly, the fluctuations are visible when testing in~$\Jtest_\Fluct$ because
\[ \big\la \u_\Fluct, \Delta \big[\w^{(1)} \big] \big\ra(x) = \frac{1}{L} \sum_{\as=1}^L 
\nabla_{\u_{\text{\rm{F}}, \as}(x)} \int_M D_{1,w_{\text{\rm{F}},\as}^{(1)}}\L(x, y)\: d\rho(y) \:. \]
Using that the first derivative of~$\ell$ vanishes in view of the EL equations,
we can write this equation in the more compact form
\[ 
\big\la \u_\Fluct, \Delta \big[\w^{(1)} \big] \big\ra(x) = \frac{1}{L} \sum_{\as=1}^L 
D_{u_{\text{\rm{F}}, \as}(x)} D_{w_{\text{\rm{F}},\as}^{(1)}} \ell(x) 
\qquad \text{for all~$x \in M$} \:. \]
These findings lead to the following definition:
\begin{Def} A jet~$\v \in (\J^1)^L$ is referred to as a {\bf{solution of the linearized field equations
with fragmentation}} if its mean~$\bar{\v}$ and fluctuation~$\v_\Fluct$ satisfy for all~$\u \in (\Jtest)^L$
and all~$x \in M$ the equations
\begin{align}
\nabla_{\bar{\u}(x)} \int_M 
\Big( \big(D_{1,\bar{v}} + \nabla_{2, \bar{\v}} \big) \L(x, y) \Big)\: d\rho(y) &= 0 \label{linmean} \\
\frac{1}{L} \sum_{\as=1}^L 
D_{u_{\text{\rm{F}}, \as}(x)} D_{v_{\text{\rm{F}},\as}} \ell(x)  &= 0\:. \label{linfluct}
\end{align}
The vector space of all linearized solutions is denoted by
\[ \Jlin = \bar{\J}^\text{\rm{lin}} \oplus \Jlin_\Fluct \;\subset\; (\J^1)^L \:. \]
\end{Def} \noindent

We point out that the linearized field equations with fragmentation
do not involve all the components of the jets, neither of the test jet~$\u$
nor of the linearized field~$\v$. Indeed, only the vector component of the
fluctuations comes into play, but their scalar component does not enter.
Moreover, if~$u_\Fluct$ is chosen as a linearized solution, then~\eqref{linfluct}
is satisfied, no matter how~$v_\Fluct$ is chosen. In other words, testing
in the direction of fluctuating linearized solutions, the equation~\eqref{linfluct}
does not give any information. Hence in the linearized field equations with fragmentation~\eqref{linmean}
and~\eqref{linfluct}, the jets~$\u$ can be changed freely in~$\Jlin_\Fluct$.

In order to implement these findings in a compact notation, it is useful to decompose the
fluctuating jets as
\[ \J^1_\Fluct = \Jc_\Fluct \oplus \Jlin_\Fluct \:, \]
where~$\Jc_\Fluct$ is a (non-orthogonal) complement of~$\Jlin_\Fluct$ in~$\J^1_\Fluct$.
We thus obtain the decomposition of the jet spaces
\[ (\J^1)^L = \bar{\J} \oplus \Jc_\Fluct \oplus \Jlin_\Fluct \:. \]
Using a block matrix notation, the unperturbed operator~$\Delta$ takes the form
\beq \label{blockform}
\la \u, \Delta \v \ra(x) = \bigg\la \begin{pmatrix} \bar{\u} \\ \u^\comp_\Fluct \\[0.2em] \u^\lin_\Fluct \end{pmatrix},
\begin{pmatrix} \bar{\Delta} & 0 & 0 \\ 0 & \Delta_\Fluct & 0 \\ 0 & 0 & 0 \end{pmatrix}
\begin{pmatrix} \bar{\v} \\ \v^\comp_\Fluct \\[0.2em] \v^\lin_\Fluct \end{pmatrix} \bigg\ra \:,
\eeq
where the operators~$\bar{\Delta}$ and~$\Delta_\Fluct$ are defined by
\begin{align}
\bar{\Delta} &: \bar{\J}^1 \rightarrow (\bJtest)^* \notag \\
&\qquad \la \bar{\u}, \bar{\Delta} \bar{\v} \ra(x) = \nabla_{\bar{\u}} \int_M 
\Big( \big(D_{1,\bar{v}} + D_{2, \bar{v}} \big) \L(x, y)+ \L(x,y)\: \bar{b}(y) \Big)\: d\rho(y) \\
\Delta_\Fluct &: \J^\comp_\Fluct \rightarrow (\J^\comp_\Fluct \cap \Jtest_\Fluct)^* \notag \\
&\qquad \la \u_\Fluct, \Delta_\Fluct v_\Fluct \ra(x) = \frac{1}{L} \sum_{\as=1}^L 
D_{u_{\text{\rm{F}}, \as}(x)} D_{v_{\text{\rm{F}},\as}} \ell(x) \:. \label{deltaFdef}
\end{align}

We finally remark that, disregarding differentiability issues,
the jet space~$\Jlin_\Fluct$ can also be understood
from the perspective of stability. If~$\rho$ is a minimizer, then
then the Hessian of~$\ell$ is non-negative and thus gives rise to
the positive semi-definite bilinear form (for details see~\cite[Section~4]{positive})
\[ \frac{1}{L} \sum_{\as=1}^L \int_M \nabla^2 \ell|_x(., .)\: d\rho \::\: (\Jtest)^L \times (\J^1)^L \rightarrow \R \:. \]
The space~$\Jlin_\Fluct$ is obtained by all fluctuating jets which are in the neutral subspace of this
positive semi-definite bilinear form.
This means that fragmentation can occur only in directions in which the Hessian of the
causal action vanishes.

\subsection{An Explicit Example} \label{secexample}
Similar to the procedure in Section~\ref{secgreen}, we may
assume that the operators~$\bar{\Delta}$ and~$\Delta_\Fluct$
in~\eqref{blockform} can be inverted by corresponding Green's operators:
\begin{Def} \label{defbarS}
A linear mapping~$\bar{S} : (\bJtest)^* \rightarrow \bar{\J}^\infty$ is referred to as an
{\bf{Green's operator for the mean}} if
\[ 
\bar{\Delta}\,\bar{S} \, \bar{\v} = -\bar{\v} \quad \text{for all~$\bar{\v} \in (\bJtest)^*$} \:. \]
A linear mapping~$S_\Fluct : (\J^\comp_\Fluct \cap \Jtest_\Fluct)^* \rightarrow \J^\comp_\Fluct
\cap \J^\infty_\Fluct$ is referred to as a
{\bf{Green's operator for fluctuations}} if
\[ 
\Delta_\Fluct \,S_\Fluct \,v_\Fluct = -v_\Fluct \quad \text{for all~$v_\Fluct \in (\J^\comp_\Fluct \cap \Jtest_\Fluct)^*$} \:. \]
\end{Def}
Before we can perform the perturbation expansion, we must
analyze how to invert the field equations on the subspace~$\Jlin_\Fluct$.
As one sees in~\eqref{blockform}, the linearized operator~$\Delta$ vanishes
on this subspace. This means that the operator on this subspace is determined
by the perturbation itself. This situation resembles the perturbation theory 
with degeneracies for the eigenvalues of a linear operator. In this case, the procedure
is to diagonalize the perturbation on the degenerate subspaces (without using
perturbation theory) before performing the perturbation expansion.
In order to explain how to proceed in our setting, we begin with a
simple concrete example.

\begin{Example} \label{exexplicit} 
{\em{Let~$\F = \R^2$ and
\[ \L\big( (x_1,x_2), (y_1, y_2) \big) = (x_1-y_1)^4 + (x_2-y_2)^2 - (x_2+y_2)^2\: (x_1-y_1)^2 \]
(for the moment, we disregard that this Lagrangian is unbounded from below; this shortcoming will
be removed after~\eqref{remove} below).
Moreover, we let~$\rho$ be the Dirac measure supported at the origin. The jet spaces are
\[ \J = \Jtest = \R \times \R^2 \;\ni\; \u = (a ,u^1, u^2) \:. \]
Obviously, all first and second partial derivatives of the Lagrangian vanish at the origin.
Therefore, $\rho$ is a critical measure, and the EL equations~\eqref{prelimb} are satisfied
for the unperturbed system with the Lagrange multiplier~$\s$ chosen to be zero.

We now consider a fragmentation with two subsystems~$L=2$, i.e.
\beq \label{JFex}
\J_\Fluct = \big\{ (\u_\as)_{\as=1,2}\:,\quad
\u_1 = -\u_2 = \big(a, u^1, u^2 \big) \in \R \times \R^2 \big\} \:.
\eeq
In order to determine~$\Jlin_\Fluct$, we first compute the Hessian of~$\ell$,
\[ \ell(x_1, x_2) = \L\big( (x_1,x_2), (0,0) \big) = x_1^4 + x_2^2 - x_2^2\, x_1^2 \:,\qquad
D^2 \ell|_{(0,0)} = \begin{pmatrix} 0 & 0 \\ 0 & 2 \end{pmatrix} \:. \]
Therefore
\beq \label{JlinFex}
\Jlin_\Fluct = \Big\{ (\u_\as)_{\as=1,2}\:,\quad
\u_1 = -\u_2 = \big(a, u^1,0 \big) \in \R \times \R \Big\} \:,
\eeq
showing that fragmentation can occur only in the $x_1$-direction.

We now prescribe the leading orders of the transformation of the universal measure~\eqref{rhotildea}
and verify if this gives a suitable starting point for a perturbative treatment.
In order to preserve the total volume, we choose~$f_\as=f_\as^{(0)}$ with
\beq \label{f0}
0 < f_1^{(0)} < 2 \qquad \text{and} \qquad f_2^{(0)} = 2-f_1^{(0)} \:.
\eeq
The transformation~$F_\as$, on the other hand, is chosen as
\beq \label{F1}
F_\as(0) = \lambda \,w_\as^{(1)}
\eeq
with the vector component
\[ w_1^{(1)} = (w,1) \qquad \text{and} \qquad w_1^{(2)} = (-w,1) \]
and~$w \in \R$.

Let us verify if this family of measures satisfies the weak EL equations,
and if not, what the resulting error is. The support of the perturbed measures consists of the two points
\beq \label{p12}
p_1 := \lambda\, (w,1) \qquad \text{and} \qquad p_2 := \lambda\, (-w,1) \:.
\eeq
Moreover, a direct computation gives
\begin{align*}
\ell(p_1) &= -8\, \lambda^4\, \big(2-f_1^{(0)} \big) \,w^2\, \big( w^2-1 \big) \\
\ell(p_2) &= -8\, \lambda^4\, f_1^{(0)}\,w^2\, \big( w^2-1 \big) \\
D\ell|_{p_1} &= -8\, \lambda^3\, \big(2-f_1^{(0)}\big) \: \Big( -w\, \big(2 w^2-1 \big),\, w^2 \Big) \\
D\ell|_{p_2} &= -8\, \lambda^3\, f_1^{(0)}\: \Big( w\, \big(2 w^2-1 \big) , w^2 \Big) \:.
\end{align*}
Hence, testing with the average and the fluctuation gives
\begin{align}
\frac{1}{2} \sum_{\as=1}^2 \nabla_{\bar{u}_\as} \ell(p_\as) &= -8\, \lambda^3\, 
\begin{pmatrix} \bar{a} \\ \bar{u}^1 \\ \bar{u}^2 \end{pmatrix}
\cdot \begin{pmatrix} \lambda\: w^2\, \big( w^2-1 \big) \\[0.2em]
\big(f_1^{(0)}-1 \big)\: w\, \big(2 w^2-1\big) \\[0.2em] w^2
\end{pmatrix} \\
\frac{1}{2} \sum_{\as=1}^2 \nabla_{(u_\Fluct)_\as} \ell(p_\as) 
 &= -8\, \lambda^3\, 
\begin{pmatrix} a \\ u^1 \\ u^2 \end{pmatrix}
\cdot \begin{pmatrix} \lambda\: \big(f_1^{(0)}-1 \big) \,w^2\, \big( w^2-1 \big) \\[0.2em] -w\, \big(2 w^2-1\big) \\[0.2em]
-\big(f_1^{(0)}-1 \big)\: w^2
\end{pmatrix} \label{nabLFex}
\end{align}
(where in the last line we parametrized the fluctuating jets as in~\eqref{JFex}).

We now restrict attention to the subspace~$\Jlin_\Fluct$ on which the unperturbed
operator~$\Delta$ in~\eqref{blockform} vanishes.
Again parametrizing according to~\eqref{JlinFex}, we obtain
\beq \label{nablapert}
\frac{1}{2} \sum_{\as=1}^2 \nabla_{(u^\lin_\Fluct)_\as}\: \ell(p_\as) = -8\, \lambda^3\, 
\begin{pmatrix} a \\ u^1 \end{pmatrix}
\cdot \begin{pmatrix} \lambda\: \big(f_1^{(0)}-1 \big) \,w^2\, \big( w^2-1 \big) \\[0.2em] -w\, \big(2 w^2-1\big)
\end{pmatrix}
\eeq
(here we simply dropped the last component in~\eqref{nabLFex}).
Moreover, the Laplacian on~$\Jlin_\Fluct$ is computed by
\begin{align}
\la &\u^\lin_\Fluct, \tilde{\Delta} \v^\lin_\Fluct\ra \notag \\
&= 4\; \left\langle \begin{pmatrix} a \\ u^1 \end{pmatrix},\;
\begin{pmatrix} 2 w^2\, (w^2-1)\: \lambda^4 & 4\, \big(f_1^{(0)}-1 \big)\, w\,(2 w^2-1)\: \lambda^3 \\[0.2em]
-w\, (2w^2-1)\:\lambda^3 & -\big(f_1^{(0)}-4 \big) (6w^2-1)\: \lambda^2 \end{pmatrix} \begin{pmatrix} b \\ v^1 \end{pmatrix} \right\rangle_{\C^2} \!\!.
\label{Lappert}
\end{align}

The basic question is whether the error in the linearized field equations~\eqref{nablapert}
can be compensated by perturbations of~$f_\as$ and~$F_\as$. Having prescribed the
leading orders by~\eqref{f0} and~\eqref{F1}, the next orders are perturbations of the form
\[ \lambda\: f^{(1)}_\as \qquad \text{and} \qquad \lambda^2 \: F^{(2)}_\as \:. \]
Substituting into~\eqref{Lappert} gives a contribution scaling like
\[ \la \u^\lin_\Fluct, \tilde{\Delta} \v^\lin_\Fluct\ra =
a\, c_5 \:\lambda^5 + u^1\: c_4\:\lambda^4 + \O(\lambda^6)\:. \]
This contribution is by a factor of~$\lambda$ {\em{smaller}} than the error
in the linearized field equations~\eqref{nablapert}.
This shows that at this stage, a perturbation expansion is not sensible.
This can be understood similar to the problem in the perturbation theory for
linear operators when applying the
naive perturbation expansion to a degenerate subspace.

The method to cure this problem is to choose~$f_1^{(0)}$ and~$w$ appropriately.
Indeed, setting
\beq \label{fwchoice}
f_1^{(0)} = 1 \qquad \text{and} \qquad w = \frac{1}{\sqrt{2}} \:,
\eeq
we obtain
\begin{align}
\frac{1}{2} \sum_{\as=1}^2 \nabla_{(u^\lin_\Fluct)_\as}\: \ell(p_\as) &= 
\begin{pmatrix} a \\ u^1 \end{pmatrix}
\cdot \begin{pmatrix} 0 \\[0.2em] 0 \end{pmatrix} \\
\la \u^\lin_\Fluct, \tilde{\Delta} \v^\lin_\Fluct\ra &= \left\langle \begin{pmatrix} a \\ u^1 \end{pmatrix},\;
\begin{pmatrix} 2 \lambda^4 & 0 \\[0.2em]
0 & 24\: \lambda^2 \end{pmatrix} \begin{pmatrix} b \\ v^1 \end{pmatrix} \right\rangle_{\C^2} \!\!.
\label{Greenex}
\end{align}
Now the linearized field equations are satisfied. This can be understood immediately by
the plot of~$\tilde{\ell}(x^1, \lambda)$ in Figure~\ref{figlplot},
\begin{figure}
\includegraphics[width=6.5cm]{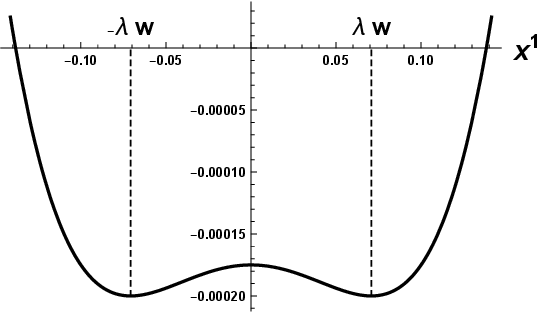}
\caption{The function~$\ell(x_1, \lambda)$ for the fragmented measure
and~$\lambda=0.1$.}
\label{figlplot}
\end{figure}%
which shows that the minima of~$\ell$ are precisely at the support points~\eqref{p12}.

Moreover, one sees that for the resulting system a perturbation expansion is sensible, provided
that the error in the linearized field equations scales like
\[ \frac{1}{2} \sum_{\as=1}^2 \nabla_{(u^\lin_\Fluct)_\as}\: \ell(p_\as) \lesssim \begin{pmatrix} a \\ u^1 \end{pmatrix}
\cdot \begin{pmatrix} \O(\lambda^5) \\[0.2em] \O(\lambda^4) \end{pmatrix} \:. \]
If this condition holds, the perturbation expansion consists in determining the jets
\[ \big( f^{(1)}_\as, F^{(2)}_\as \big) \:,\quad \big( f^{(2)}_\as, F^{(3)}_\as \big)
\:,\quad \big( f^{(3)}_\as, F^{(4)}_\as \big) \:,\ldots \]
iteratively. For example, we may modify the Lagrangian by adding a potential of sixth order
\beq \label{remove}
\L\big( (x_1,x_2), (y_1, y_2) \big) \;\rightarrow\;
\L\big( (x_1,x_2), (y_1, y_2) \big) + \big(x_1^6 + x_2^6 + y_1^6 + y_2^6 \big) \:.
\eeq
After this modification, the Lagrangian is bounded below.
By adding an irrelevant constant, it can even be arranged to be non-negative.
}} \QEDrem
\end{Example}

\subsection{The Perturbation Expansion} \label{secpertfluct}
After these preparations, we now give the general construction.
We choose the unperturbed scalar components such as to preserve the total volume, i.e.\
\beq \label{f0choice}
f^{(0)}_\as \geq 0 \qquad \text{and} \qquad \frac{1}{L} \sum_{\as=1}^L f^{(0)}_\as = 1 \:.
\eeq
Next, we choose the linearized solution which triggers the fragmentation.
In order to allow for a more general scaling, we make the ansatz
\beq \label{F1choice}
\w^{(1)} = \lambda^p\: \bar{\v}^\lin+ \lambda^q\: v^\lin_\Fluct
\eeq
with parameters~$p,q>0$ and
\[ \min(p,q)=1 \]
(here~$v^\lin_\Fluct$ denotes a jet with vanishing scalar component).

For the perturbation expansion, we again work with the function~$c$ defined by~\eqref{bdef}
and expand according to~\eqref{Fxser} and~\eqref{bser}. We also again use the
notation~\eqref{vpdef} (but of course, now all objects carry additional subsystem indices~$\as$
or~$\bs$). Our ansatz~\eqref{f0choice} and~\eqref{F1choice}
means that the following jets are already determined:
\[ c^{(0)}_\as(x) = \log f^{(0)}_\as(x) \:,\qquad F^{(0)}_\as(x)=x \]
and
\[ \left\{ \begin{array}{ll} \w^{(p)} = \bar{\v}^\lin + v^\lin_\Fluct & \text{if~$p=q$} \\[0.3em]
\w^{(p)} = \bar{\v}^\lin \quad \text{and }  \quad \w^{(q)} = v^\lin_\Fluct & \text{if~$p \neq q \:.$}
\end{array} \right. \]
We set all other jets~$\w^{(\ell)}$ to zero.
Now we proceed in two steps. We first perturb only in the jet spaces~$\bar{\J} \oplus \J^\comp_\Fluct$,
beginning to the order~$p+1$. Thus we modify the jets according to
\[ \left\{ \begin{array}{ll} 
\w^{(n)} = \bar{\w}^{(n)} + (\w^\comp_\Fluct)^{(n)} & \text{if~$n>p$, $n \neq q$} \\
\w^{(n)} = v^\lin_\Fluct + \bar{\w}^{(n)} + (\w^\comp_\Fluct)^{(n)} & \text{if~$n>p$, $n = q \:.$}
\end{array} \right. \]
The jets~$\bar{\w}^{(n)}$ and~$(\w^\comp_\Fluct)^{(n)}$ can be computed iteratively
for~$n=p+1,p+2, \ldots$ by multiplying the error in the weak EL equations
by the Green's operators in Definition~\ref{defbarS}.
We let~$\tilde{\rho}$ be the measure obtained from this perturbation expansion
according to~\eqref{rhotildea}. By construction, this measure satisfies
the weak EL equations~\eqref{prelimb} if tested in the direction of~$\bar{\J} \oplus \J^\comp_\Fluct$, i.e.\
\[ \frac{1}{L} \sum_{\as=1}^L \nabla_{\u_\as} \tilde{\ell} \big(F_\as(x) \big) = 0 \qquad
\text{for all~$\u \in (\bar{J} \oplus \J^\comp_\Fluct) \cap \Jtest$} \:, \]
where the tilde refers to the perturbed measure,
\[ \tilde{\ell}(F_\as(x)) := \frac{1}{L} \sum_{\bs=1}^L \int_M \L\big(F_\as(x), F_\bs(y) \big)\: f_\bs(y)\: d\rho(y) -\s \:. \]
However, the weak EL equations will not hold in general if we test in the direction of~$\Jlin_\Fluct$.
In order to obtain a well-defined perturbation expansion, we need to assume that
the error in the EL equation is small compared to the size of the Laplacian
on~$\Jlin_\Fluct$, as is made precise in the following definition.
\begin{Def} \label{defwellposed}
The Laplacian on~$\Jlin_\Fluct$ is {\bf{definite of order~$r$}} if
there is an operator~$T_\Fluct : (\Jlin_\Fluct \cap \Jtest_\Fluct)^* \rightarrow \Jlin_\Fluct$ with the property that
for all~$\u \in \Jlin_\Fluct \cap \Jtest$ and~$\v \in (\Jlin_\Fluct \cap \Jtest_\Fluct)^*$,
\beq \label{defr}
\Big\la \big(a^\lin_\Fluct, \lambda^q \, u^\lin_\Fluct\big), \tilde{\Delta} \,T_\Fluct\,
\big(b^\lin_\Fluct, \lambda^q \, v^\lin_\Fluct \big) \Big\ra 
= \lambda^r \: \big\la \u^\lin_\Fluct, \v^\lin_\Fluct \big\ra \Big( 1 + \O(\lambda) \Big) \:.
\eeq
The ansatz~\eqref{f0choice} and~\eqref{F1choice} gives rise to a  {\bf{well-posed
fragmentation}} if there is~$r>q$ such that the Laplacian on~$\Jlin_\Fluct$ is definite of order~$r$ and
if for all~$\u \in \Jlin_\Fluct \cap \Jtest$,
\beq \label{flucterr}
\frac{1}{L} \sum_{\as=1}^L \Big( \big(a^\lin_\Fluct \big)_\as(x) + \lambda^q \, D_{\big(u^\lin_\Fluct \big)_\as} \Big)
\tilde{\ell}\big(F_\as(x) \big) = \O\big( \lambda^{r+1} \big) \:.
\eeq
\end{Def}
If the condition in this definition holds, the weak EL equations can also be satisfied in
the direction of~$\Jlin_\Fluct$ by changing the perturbation ansatz according to
\beq \label{addjet}
\tilde{\w} \rightarrow \tilde{\w}
+ \lambda \, \Big(\big(c^\lin_\Fluct\big)^{(1)}, \lambda^q \,\big(w^\lin_\Fluct\big)^{(q+1)} \Big) 
+ \lambda^2 \, \Big(\big(c^\lin_\Fluct\big)^{(2)}, \lambda^q \,\big(w^\lin_\Fluct\big)^{(q+2)} \Big)  + \cdots \:.
\eeq
Now the error in the EL equations in the direction of~$\Jlin_\Fluct$ can be compensated
order by order by multiplying with the corresponding Green's operator~$T_\Fluct$.
Clearly, the higher order jets~$\bar{\w}^{(n)}$ and~$(\w^\comp_\Fluct)^{(n)}$ are
also affected by the jets added in~\eqref{addjet}, but the resulting error can be
compensated again using the Green's operators in Definition~\ref{defbarS}.
In this way, we obtain a perturbation expansion for the universal measure
with fragmentation. The expansion is well-defined as a formal power series in~$\lambda$.

We note that the different fragments of the measure are separated by~$\tilde{w}^\lin_\Fluct \sim \lambda^q$.
Therefore, the ``size'' of the microstructure obtained by fragmentation is of order~$\sim \lambda^q$.
Consequently, differentiating this microstructure gives a scaling factor~$\lambda^{-q}$.
This is the reason why on the left side of~\eqref{defr} and~\eqref{flucterr},
the vector components of the jets are multiplied by scaling factors~$\lambda^q$.
In Example~\eqref{exexplicit}, this scaling behavior can be seen explicitly
from the different powers of~$\lambda$ in~\eqref{Greenex}.

For clarity, we also point out that in the applications, the delicate step is to
choose the weights~$f_\as^{(0)}$ as well as the ansatz~\eqref{F1choice}
correctly such as to obtain a well-posed fragmentation. This difficulty
already became clear in Example~\ref{exexplicit}, where we had to
come up with the ansatz~\eqref{fwchoice} and choose~$p=q$,
giving a well-posed fragmentation with~$r=4$.
Once the correct ansatz for the fragmentation has been found,
the perturbation expansion can be performed in a straightforward way
as outlined above. We postpone the combinatorial details to the physical
applications in~\cite{qftlimit}.

\section{Perturbation Expansion for Causal Fermion Systems} \label{seccfs}
\subsection{Preliminaries} \label{seccfsprelim}
We briefly recall how the causal action principle for causal fermion systems
fits into the framework of causal variational principles in the non-compact setting
(see also~\cite[Section~2.3]{jet}).
Compared to the setting in Section~\ref{secmotivate} 
and~\cite[Section~1.1]{cfs}, we incorporate
the trace constraint by restricting attention to operators of fixed trace.
Moreover, we treat the boundedness constraint with a Lagrange multiplier~$\kappa$.
Finally, by assuming that the unperturbed measure has the property that all space-time
points are regular (see~\cite[Definition~1.1.5]{cfs}), we may assume that all operators
have exactly~$n$ positive and~$n$ negative eigenvalues. This leads to the following setting:

Let~$(\H, \la .|. \ra_\H)$ be a complex Hilbert space. Moreover, we are given
parameters~$n \in \N$ (the spin dimension), $c > 0$ (the constraint for the local trace)
and~$\kappa>0$ (the Lagrange multiplier of the boundedness constraint).
We let~$\F \subset \Lin(\H)$ be the set of all operators on~$\H$ with the following properties:
\begin{itemize}[leftmargin=2em]
\itemD $F$ is selfadjoint, has finite rank and (counting multiplicities) has~$n$ positive and~$n$ negative eigenvalues. \\[-0.8em]
\itemD The trace is constant, i.e.
\beq \label{loctrace}
\tr(F) = c\:.
\eeq
\end{itemize}
On~$\F$ we consider the topology induced by the sup-norm on~$\Lin(\H)$.
If~$\H$ is finite-dimensional, then~$\F$ has a smooth manifold structure
(see the concept of a flag manifold in~\cite{helgason} or the detailed construction in~\cite[Section~2.4]{intro}).

We introduce the Lagrangian~$\L_\kappa$ by adding a Lagrange multiplier term to~\eqref{Lagrange},
\beq \label{Lkappa}
\L_\kappa \::\: \F \times \F \rightarrow \R\:,\qquad
\L_\kappa(x,y) = \big| (xy)^2 \big| - \frac{1}{2n}\: |xy|^2 + \kappa\: |xy|^2 \:.
\eeq
Clearly, this Lagrangian is non-negative and continuous on~$\F \times \F$. Thus we are back in the setting of
Section~\ref{secnoncompact}. The EL equations in Definition~\ref{defcritical} agree with
the EL equations as derived for the causal action principle with constraints
in~\cite{lagrange} (see~\cite[Theorem~1.1]{lagrange}).

Before going on, we make a few remarks. Since in the present setting,
the Lagrange multiplier term~$\kappa\, |xy|^2$ in~\eqref{Lkappa} is always present,
we can simplify the notation by always omitting the subscript~$\kappa$.
We also point out that we shall always keep the constants~$c$ and~$\kappa$ in~\eqref{loctrace}
and~\eqref{Lkappa} fixed when varying or perturbing the measure~$\rho$.
This is justified as follows. The constant~$c$ can be changed arbitrarily by rescaling the measure
according to
\[ \rho(\Omega) \rightarrow \rho\Big( \big\{ \alpha x \:\big|\: x \in \Omega \big\} \Big) \qquad \text{with} \qquad \alpha \in \R\:. \]
Combining this transformation with our previous transformation~\eqref{rescale1}, the freedom
in rescaling the universal measure is exhausted. Therefore, the parameter~$\kappa$ must
be regarded as a physical parameter of the system. The reason for keeping it fixed is that
we want to describe {\em{localized physical systems}}, meaning that the perturbations of~$\rho$ are
spatially compact or that the resulting space-time is asymptotically flat. In such situations, the parameter~$\kappa$
is determined by the asymptotic form of the universal measure at infinity, which is kept fixed
in our variations and perturbations.
More generally, $\kappa$ can be kept fixed if we assume that there is a macroscopic region
in space-time where no interaction takes place.

We now recall the definition of a few other basic objects needed for the analysis of causal
fermion systems (for more details see~\cite[Section~1.1]{cfs}).
For every~$x \in \F$ we define the {\em{spin space}}~$S_x$ by~$S_x = x(\H)$; it is a subspace
of~$\H$ of dimension~$2n$. On the spin space~$S_x$, the {\em{spin scalar product}} $\Sl .|. \Sr_x$
is defined by
\[ \Sl u | v \Sr_x = -\la u | x u \ra_\H \qquad \text{(for all $u,v \in S_x$)}\:. \]
We let~$\pi_x$ be the orthogonal projection on~$S_x \subset \H$.
Then, for any~$x, y \in M$ we define the {\em{kernel of the fermionic projector}}~$P(x,y)$ by
\[ P(x,y) = \pi_x \,y|_{S_y} \::\: S_y \rightarrow S_x \:. \]
The kernel of the fermionic projector is very useful because, forming the closed chain~$A_{xy}$ by
\[ A_{xy} := P(x,y)\, P(y,x) = \pi_x \,y\, x|_{S_x}\::\: S_x \rightarrow S_x\:, \]
the eigenvalues of~$A_{xy}$ coincide with the
eigenvalues~$\lambda^{xy}_1, \ldots, \lambda^{xy}_{2n}$ in~\eqref{sw}.
In this way, the Lagrangian can be expressed in terms of
the kernel of the fermionic projector.

A {\em{wave function}}~$\psi$
is defined as a mapping which to every~$x \in M$ associates a vector of the corresponding spin space,
\[ \psi \::\: M \rightarrow \H \qquad \text{with} \qquad \psi(x) \in S_xM \quad \text{for all~$x \in M$}\:. \]
A wave function is said to be {\em{continuous}} at~$x$ if
for every~$x \in M$ and~$\varepsilon>0$, there is~$\delta>0$ such that
\[ \big\| \sqrt{|y|} \,\psi(y) -  \sqrt{|x|}\, \psi(x) \big\|_\H < \varepsilon
\qquad \text{for all~$y \in M$ with~$\|y-x\| \leq \delta$} \:. \]
The vector space of continuous wave functions is denoted by~$C^0(M, SM)$.
For every~$u \in \H$, the corresponding {\em{physical wave function}}~$\psi^u$
is the wave function obtained by projecting to the spin spaces, i.e.
\[ \psi^u(x) := \pi_x u \in S_xM \:. \]
The physical wave functions can be understood as describing the 
``occupied states'' of the system (for details see~\cite[\S1.1.4 and~\S1.2.4]{cfs}).
The physical wave functions are all continuous. The
{\em{wave evaluation operator}}~$\Psi$ is the linear operator which 
to every Hilbert space vector associates the corresponding physical wave function,
\beq \label{weo}
\Psi \::\: \H \rightarrow C^0(M, SM)\:, \qquad u \mapsto \psi^u \:.
\eeq
Evaluating at a fixed space-time point gives the mapping
\[ \Psi(x) \::\: \H \rightarrow S_xM\:, \qquad u \mapsto \psi^u(x) \:. \]
The operator~$x$ as well as the
kernel of the fermionic projector can be expressed in terms of the wave evaluation operator by
(see~\cite[Lemma~1.1.3]{cfs})
\beq \label{xPid}
x = - \Psi(x)^* \,\Psi(x) \qquad \text{and} \qquad P(x,y) = -\Psi(x)\, \Psi(y)^*\:.
\eeq

\subsection{Perturbation Expansion for the Wave Evaluation Operator} \label{secweo}
The perturbation expansion in Section~\ref{secabstract} was performed in a chart
on~$\F$. We now explain how to construct such a chart. Working in this chart
will also immediately give a perturbation expansion for the wave evaluation operator.
Given~$x \in M$, we consider the mapping
\beq \label{phidef}
R \::\: \big\{ \psi \in \Lin(\H, S_x) \:\big|\:  \tr (\psi^* \psi) \neq 0 \big\}
\rightarrow \Lin(\H) \:, \qquad \psi \mapsto \frac{c}{\tr (\psi^* \psi)}\: \psi^* \psi \:.
\eeq
The operators in the image of~$R$ are selfadjoint, have finite rank and at most~$n$
positive and at most~$n$ negative eigenvalues. Moreover,
due to the rescaling by the prefactor~$c/\tr (\psi^* \psi)$, they satisfy the trace condition~\eqref{loctrace}.
Let us verify that the image of~$R$ contains all operators in~$\F$:
By definition of~$\F$, a given operator~$F \in \F$ is selfadjoint and has~$n$ positive
eigenvalues (which we denote by~$\nu_1, \ldots, \nu_n>0$) and~$n$ negative eigenvalues
(denoted by~$(\nu_{n+1}, \ldots, \nu_{2n})$).
Diagonalizing~$F$ gives a representation
\[ F = U \: \text{diag}(\nu_1, \ldots, \nu_{2n}) \: U^* \]
where~$U : \C^{2n} \rightarrow \H$ is an isometric embedding. It is useful to rewrite this
equation as
\[ F = V^* \: \text{diag}(\underbrace{1, \ldots, 1}_{\text{$n$ entries}}, \underbrace{-1, \ldots, -1}_{\text{$n$ entries}}) \: V \]
with~$V := \text{diag}(\sqrt{|\nu_1|}, \ldots, \sqrt{|\nu_{2n}|}) \, U^*$.
Then, choosing a pseudo-orthogonal basis $(e_\alpha)_{\alpha=1,\ldots, 2n}$ of~$S_x$, the mapping
\[ \psi : \H \rightarrow S_x \:, \qquad \psi (u) := \sum_{\alpha=1}^{2n} (V u)^\alpha\: e_\alpha \]
has the desired property~$F = \psi^* \psi$.

But the mapping~$R$ is not injective for two reasons: First, due to the rescaling, multiplying~$\psi$ by a complex number leaves~$R(\psi)$ unchanged.
Second, a local unitary transformation
\beq \label{locunitary}
\psi \rightarrow U\,\psi  \qquad \text{with} \qquad U \in \U(S_x)
\eeq
preserves the combination~$\psi^* \psi$ and thus leaves~$R(\psi)$ unchanged.

The mapping~$R$ can be used to construct a chart of~$\F$ around~$x$:
Since the image of~$R$ contains~$\F$, the operator~$x$ can be written as~$x = R(\psi)$ with~$\psi \in \Lin(\H, S_x)$
(more explicitly, we can choose~$\psi = \Psi(x)$).
By continuity, the numbers of positive and negative eigenvalues
of the operator~$R(\phi)$ are again equal to~$n$ 
for all~$\phi$ in a small neighborhood~$V \subset \Lin(\H, S_x)$ of~$\psi$.
Thus the restriction of~$R$ to this neighborhood maps to~$\F$,
\[ R|_V \::\: \Lin(\H, S_x) \cap V \rightarrow \F \:. \]
Differentiating at~$\psi$ gives a linear operator~$DR|_\psi : \Lin(\H, S_x) \rightarrow T_x\F$.
This operator is not injective (because infinitesimal scalings and unitary transformations~\eqref{locunitary}
lie in its kernel). Therefore, we choose a proper subspace~$E \subset \Lin(\H, S_x)$ such that
the restriction to~$E$ is invertible,
\[ DR|_\psi \big|_E \::\: E \subset \Lin(\H, S_x) \rightarrow T_x\F \qquad \text{is continuously invertible} \]
(if~$\H$ is finite-dimensional, such a subspace~$E$ always exist; in the infinite-dimensional setting the
condition that the inverse be continuous poses constraints which we shall not analyze here).
As a consequence, the restriction of~$R$ is a local diffeomorphism, meaning that
there is an open neighborhood~$V' \subset V \subset \Lin(\H, S_x)$ of~$\psi$ 
and an open neighborhood~$U \subset \F$ of~$x$ such that the restriction
\[ R|_{\psi + (E \cap V')} \::\: \psi + (E \cap V') \rightarrow  U \subset \F \]
is a diffeomorphism (here~$\psi + E$ denotes the affine subspace through~$\psi$). Its inverse
\[ X := (R|_{\psi + (E \cap V')})^{-1} \::\: U \subset \F \rightarrow \psi + E \]
defines a chart~$(X,U)$ around~$x$.
Choosing a basis~$(e_1, \ldots, e_m)$ of~$E$, we write the mapping~$F : M \rightarrow \F$ in
components~$F(x)^\alpha$, i.e.\
\[ X\big(F(x) \big) = \psi + \sum_{\alpha=1}^m F(x)^\alpha \:e_\alpha \:. \]

Choosing for every~$x \in M$ a chart of this form and choosing a suitable jet space~$\Jtest$,
we are back in the setting of Section~\ref{secrhopert}.
After determining the~$F^{(p)}$, the corresponding perturbation of the wave evaluation
operator is given simply by the component in our chart, i.e.
\[ \Psi^{(p)}(x) = F^{(p)}(x)^\alpha \: e_\alpha \;\in\; E \subset \Lin(\H, S_x)  \qquad (p \geq 1)\:. \]

\subsection{Perturbing the Vacuum} \label{seccfsvacpert}
We now explain how the construction in Section~\ref{secvacpert} applies to causal fermion systems.
Let~$\rho$ be a universal measure describing the vacuum (for example, a regularized Dirac sea
configuration as constructed in~\cite[Section~1.2]{cfs}).
Introducing particles and/or anti-particles
(as described in~\cite[Section~\S2.1.7]{cfs}) amounts to modifying the wave evaluation operator~$\Psi$ to
\beq \label{hatPsi}
\hat{\Psi} := \Psi + \Delta \Psi \::\: \H \rightarrow C^0(M, SM)\:.
\eeq
At this point, the complication arises that the local correlation operators defined 
in analogy to~\eqref{xPid} by~$\hat{F}(x) = -\hat{\Psi}(x)^* \hat{\Psi}(x)$ (see~\cite[eq.~(1.4.12)]{cfs})
will in general violate our trace condition~\eqref{loctrace}. In order to resolve
this problem, we rescale the local correlation operators similar as in~\eqref{phidef} by setting
\beq \label{rescale}
\hat{H}(x) := \frac{c}{\tr \big( \hat{\Psi}(x)^* \hat{\Psi}(x) \big)}\: \hat{\Psi}(x)^* \hat{\Psi}(x) \:.
\eeq
We now introduce the corresponding universal measure~$\hat{\rho}$ as the push-forward of~$\hat{H}$,
\beq \label{hatrhoH}
\hat{\rho} := \hat{H}_* \rho \:.
\eeq
Now we are back in the setting of Section~\ref{secvacpert}.
We remark that the rescaling~\eqref{rescale} seems unproblematic because in physical applications
it affects only the higher orders in~$\varepsilon$ relative to the length scale of macroscopic
physics (for details on this point see~\cite[Section~2.5]{cfs}).

\section{Example: Perturbation Expansion in the Continuum Limit} \label{seccontinuum}

\subsection{Preliminaries} We now recall a few constructions of the continuum limit analysis in~\cite{cfs}
which will be of relevance here.
In~\cite[\S1.4.1]{cfs} the EL equations are written in a form which is particularly convenient
for a detailed analysis. These EL equations are obtained by considering a special class of
variations of the wave evaluation operator~$\Psi$:
\begin{Def} \label{defvarc}
A {\bf{variation of the physical wave functions}}~$(\Psi_\tau)_{\tau \in (-\tau_{\max}, \tau_{\max})}$
with~$\tau_{\max}>0$ and~$\Psi_0=\Psi$ is {\bf{smooth}} and {\bf{compact}} if the family of operators
has the following properties:
\begin{itemize}[leftmargin=2em]
\item[\rm{(a)}] The variation is trivial on the orthogonal complement of a finite-dimensional
subspace~$I \subset \H$, i.e.
\[ \Psi_\tau |_{I^\perp} = \Psi \qquad \text{for all~$\tau \in (-\tau_{\max}, \tau_{\max})$} \:. \]
\item[\rm{(b)}] There is a compact subset~$K \subset M$ outside which the variation is trivial, i.e.
\[ \big( \Psi_\tau(u) \big) \big|_{M \setminus K} = \big( \Psi(u) \big) \big|_{M \setminus K}
\qquad \text{for all~$\tau \in (-\tau_{\max}, \tau_{\max})$ and~$u \in \H$} \:. \]
\item[\rm{(c)}] The Lagrangian is continuously differentiable in the sense that the derivative
\[ \frac{d}{d\tau} \L\big( x, F_\tau(y) \big) \big|_{\tau=0} \qquad \text{with} \qquad
F_\tau(x) := \frac{c}{\tr (\Psi_\tau(x)^* \Psi_\tau(x))}\:\Psi_\tau(x)^* \Psi_\tau(x) \]
exists and is continuous on~$M \times M$.
\end{itemize}
\end{Def}

For clarity, we point out that, similar to~\eqref{phidef},
the factor~$c/\tr (\Psi_\tau(x)^* \Psi_\tau(x))$ is again needed in order to
the trace condition~\eqref{loctrace}. For the derivation of the EL equations, it is more
convenient to disregard this condition in the variation, and to realize
it instead by a Lagrange multiplier term. 
Then, according to~\eqref{xPid}, the first variation~$\delta \Psi = \partial_\tau \Psi|_{\tau=0}$
defines a corresponding variation of the kernel of the fermionic projector given by
\beq \label{delP}
\delta P(x,y) = -\delta \Psi(x)\, \Psi(y)^* -\Psi(x)\, \delta \Psi(y)^* \:.
\eeq
The resulting first variation of the Lagrangian can be written as (see~\cite[Section~5.2]{pfp}
and~\cite[eq.~(1.4.16)]{cfs})
\beq \label{delLdef}
\delta \L(x,y) = 
\Tr_{S_y} \big( Q(y,x)\, \delta P(x,y) \big) + \Tr_{S_x} \!\big( Q(x,y)\, \delta P(x,y)^* \big)
\eeq
with a kernel~$Q(x,y) : S_y \rightarrow S_x$ which is symmetric in the sense that
\[ Q(x,y)^* = Q(y,x) \]
(a more explicit formula for~$Q(x,y)$ is given in~\cite[Lemma~3.6.2]{cfs}).
Then the EL equations corresponding to the above variations can be written
as (see~\cite[Proposition~1.4.3]{cfs})
\beq \label{Qrel}
\int_M Q(x,y)\, \psi^u(y)\: d\rho(y) = \frac{\lambda}{2}\: \psi^u(x) \qquad \text{for all~$u \in \H$ and~$x \in M$}\:,
\eeq
where~$\lambda$ is the Lagrange multiplier needed in order to arrange the
trace condition~\eqref{loctrace}.
The connection to the weak EL equations~\eqref{ELtest} is not obvious and will be
explained in Section~\ref{secchoosejet} below.

In the {\em{continuum limit}} (for details see~\cite[\S3.5.2]{cfs}), the EL equations~\eqref{Qrel} are
evaluated for a physical wave function~$\psi^u$ having the form of an
{\em{ultrarelativistic wave packet}} of negative energy, meaning that the
wave packet has frequency of the order~$|\Omega|$ and is spatially localized
on the scale~$\delta$ (as measured in a chosen reference frame). Moreover, we assume that the spatial distance
 of the ultrarelativistic wave packet from the space-time point~$x$ is on the scale~$\ell$ with
(see~\cite[eq.~(3.5.28) and Figure~3.1]{cfs})
\beq \label{s:scales}
\varepsilon \ll |\Omega|^{-1} \ll \delta \ll \ell, \ell_\text{macro}, m^{-1}
\eeq
(where~$m^{-1}$ is the Compton scale and~$\ell_\text{macro}$ denotes the
length scales of atomic or high energy physics).
Moreover, the equations~\eqref{Qrel} are evaluated weakly with a test
function~$\phi$ which is supported in a $\delta$-neighborhood of the point~$x$
(with Euclidean distances measured again in a chosen reference frame).
Then the supports of~$\phi$ and~$\psi^u$ are disjoint, so that the right side
of~\eqref{Qrel} vanishes (see~\cite[eqs~(3.5.24) and~(3.5.29)]{cfs})
\beq \label{Qrelweak}
\int_M d\rho(x) \int_M d\rho(y)\: \Sl \phi(x) \,|\, Q(x,y)\, \psi^u(y) \Sr_x = 0\:.
\eeq
Written in this form, the main contribution to the EL equations comes from the
behavior of~$Q(x,y)$ on the light cone, making it possible to analyze the equations
in detail in the formalism of the continuum limit
(for details see~\cite[Section~2.4 and Chapters~3-5]{cfs}).

In the resulting continuum description, the kernel of the fermionic projector is a solution of the Dirac equation
in the presence of a classical gauge field.
In order to keep the setting as simple as possible, we here restrict attention to 
one type of elementary particles and a $\U(1)$ gauge field
(the generalizations to several generations and more general gauge fields
are carried out in detail in~\cite[Chapters~3--5]{cfs}). Then the Dirac equation reads
\beq 
\big(i \Pdd + \slashed{A} - m \big) \, P(x,y) = 0\:, \label{s:diracPaux}
\eeq
where~$A$ can be thought of as an electromagnetic potential, but it does not need to satisfy
Maxwell's equations.
In order to construct the kernel of the fermionic projector in the presence of the electromagnetic potential, one
expands the Dirac equation~\eqref{s:diracPaux} in powers of the potential
and solves the equations iteratively with the help of Dirac Green's operators~$s$ defined by
\beq \label{diracgreen}
(i \Pdd - m)\, s_m(x,y) = \delta^4(x-y) \:.
\eeq
The resulting {\em{causal perturbation expansion}} becomes unique by
making use of the underlying causal structure (for details see~\cite[Section~2.1]{cfs}).

\subsection{Choosing the Jet Spaces and the Green's Operator} \label{secchoosejet}
In this section we explain how the weak EL equations~\eqref{ELtest} and the
perturbation expansion of Section~\ref{secabstract} are related to the analysis in the continuum limit.
Our first task is to introduce the jet spaces. It is useful that, similar as explained in~\eqref{hatPsi} and~\eqref{rescale}
for finite variations, tangent vectors to~$\F$ on~$M$ 
can be described by infinitesimal variations of the wave evaluation operator.
Thus we describe a tangent vector~$u \in T_x\F$ at a space-time point~$x \in M$ as
\beq \label{varyPsi}
u = \delta \hat{H}(x) = 
-\delta \Psi(x)^*\, \Psi(x) - \Psi(x)^*\, \delta \Psi(x)
+ \frac{x}{c}\: \tr \big( \delta \Psi(x)^*\, \Psi(x) + \Psi(x)^*\, \delta \Psi(x)\big)
\eeq
(where we used that~$\tr x =-\tr \Psi(x)^* \Psi(x) = c$) with
\beq \label{deltaPsi}
\delta \Psi \::\: \H \rightarrow C^\infty(M, SM)\:.
\eeq

Our next goal is to introduce the space of test jets~$\Jtest$ in such a way that the 
weak EL equations~\eqref{ELtest} agree with the EL equations in the continuum limit~\eqref{Qrelweak}
for~$\psi^u$ an ultrarelativistic wave packet~\eqref{s:scales}.
We say that a physical wave function~$\psi^u$ is {\em{macroscopic}} if its energy and momentum
is much smaller than the Planck energy. We choose~$u$ such that~$\psi^u$ is macroscopic
and is an {\em{ultrarelativistic wave packet}} as defined before~\eqref{s:scales}.
Next, we choose~$\delta \psi^u$ as a wave function with compact support such that
its spatial distance to the ultrarelativistic wave packet
scales like
\beq \label{scalenew}
\varepsilon \ll \text{dist} \big( \supp \delta \psi^u, \supp \psi^u \big) \ll \ell_\text{macro} \:.
\eeq
We define the corresponding variation of the wave evaluation operator~$\delta \Psi$
as the unique linear mapping with the properties that
\[ \delta \Psi \::\: v \mapsto \left\{ \begin{array}{cl}
\delta \psi^u & \text{if~$v=u$} \\
0 & \text{if~$v \perp u\:.$} \end{array} \right. \]
Since by construction, $\psi^u$ and~$\delta \psi^u$ have disjoint supports,
the trace in~\eqref{varyPsi} vanishes. Therefore, the vector field
described by~$\delta \Psi$ is given by
\[ u = \delta \hat{H}(x) = 
-\delta \Psi(x)^*\, \Psi(x) - \Psi(x)^*\, \delta \Psi(x) \:. \]
We choose~$\Gtest$ as the span of all the vector fields~$u$ for~$\delta \Psi$
as specified above. Since in the weak evaluation on the light cone, only variations of the wave functions are considered,
we choose the scalar component of~$\Jtest$ trivially,
\beq \label{Jtestcl}
\Jtest = \{0\} \oplus \Gtest \;\subset\; C^\infty(M, \R) \oplus C^\infty(M, T\F) \:.
\eeq
We remark that there is no point in making~\eqref{scalenew} mathematically more precise, because
in the formalism of the continuum limit one also works merely with the scaling behavior.

The next lemma gives the connection between the weak EL equations~\eqref{ELtest} and
their continuum limit~\eqref{Qrelweak}.
\begin{Lemma} For any~$\u \in \Jtest$ and all~$x \in M$,
\[ \nabla_\u \ell(x) = -2 \re \int_M \tr \big( \delta \Psi(x)^* \, Q(x,y)\, \Psi(y) \big)\: d\rho(y) \:. \]
\end{Lemma}
\Proof Since~$\u$ has no scalar component, the term involving~$\s$ in~\eqref{ldef} drops out.
Using~\eqref{delP} together with the fact that the jet~$\u$ acts only on~$x$,
\[ \nabla_{\u(x)} P(x,y) = -\delta \Psi(x)\, \Psi(y)^* \:. \]
Using this formula in~\eqref{delLdef}, we obtain
\begin{align*}
\nabla_{\u(x)}  \L(x,y) &= 
-\Tr_{S_y} \big( Q(y,x)\, \delta \Psi(x)\, \Psi(y)^* \big) - \Tr_{S_x} \!\big( Q(x,y)\, \Psi(y)\, \delta \Psi(x)^* \big) \\
&= -2 \, \re \tr \big( \delta \Psi(x)^*\, Q(x,y)\, \Psi(y) \big) \:,
\end{align*}
where in the last step we cyclically commuted the factors inside the trace.
Integrating over~$y$ gives the result.
\QED

We next turn attention to the jets used for perturbing the measure.
The abstract Definition~\ref{defJvary} is intended to make~$\J^\infty$ as large
as possible, giving the largest possible freedom for the perturbations.
But not all of the degrees of freedom of~$\J^\infty$ are needed in the applications.
Therefore, we must specify those subspaces of~$\J^\infty$ which are of relevance here.
We first consider jets which are needed to describe particle
and anti-particle states.
\begin{Def} A vector field~$u$ of the form~\eqref{varyPsi} where the variation~$\delta \Psi$
is a mapping of finite rank with the property that
for every~$u \in \H$, either~$\Psi u$ or~$\delta \Psi u$ is macroscopic,
is called {\bf{fermionic vector field}}. The vector space of fermionic vector fields
is referred to as~$\Gfermi$. The {\bf{fermionic jets}} are defined by
\[ \Jfermi = \{0\} \oplus \Gfermi \:. \]
\end{Def} \noindent
In the next definition we introduce the jets describing the bosons,
for simplicity for an electromagnetic potential.
\begin{Def} Let~$A \in C^\infty(M, T^*M)$ be a smooth one-form.
A vector field~$u$ of the form~\eqref{varyPsi} with
\beq \label{Bjet}
\big(\delta \Psi\big)(x)= -\int_M s_m(x,y) \, \slashed{A}(y)\, \Psi(y)\: d\rho(y)
\eeq
is called {\bf{bosonic vector field}} (here~$s_m(x,y)$ is a Dirac Green's function~\eqref{diracgreen}).
The vector space of bosonic vector fields
is referred to as~$\Gbose$. The {\bf{bosonic jets}} are defined by
\[ \Jbose = \{0\} \oplus \Gbose \:. \]
\end{Def} \noindent
Clearly, the fermionic and bosonic jets are subspaces of~$\J^\infty$,
\[ \Jfermi, \Jbose \subset \J^\infty \:. \]

We now explain how the perturbative description in the continuum limit
is described in our setting. In the formalism of Section~\ref{secvacpert},
the particles and anti-particles as introduced in~\cite[\S3.4.3]{cfs}
correspond to a perturbation~$H$ of the vacuum measure in~\eqref{rhohat}.
The corresponding jets in~\eqref{wpex} are fermionic,
\[ \v^{(p)} \in \Jfermi \:. \]
The resulting contributions to the weak EL equations are compensated by
bosonic fields. Consequently, we here introduce the Green's operator~$S$
(see Definition~\ref{defS}) as a mapping to the bosonic jets,
\beq \label{Sboson}
S \::\: (\Jtest)^* \rightarrow \Jbose \subset \J^\infty \:.
\eeq
The condition~\eqref{Sdefine} means that the potential~$\B$ in~\eqref{Bjet}
satisfies the inhomogeneous classical field equations. In the example of
an electromagnetic potential~\eqref{s:diracPaux}
a Maxwell field, these equations become
\[ 
\partial_{jk} (S \v)^k - \Box (S \v)_j = -c \,v_j \]
(or equivalently with differential forms~$\delta d \,S\, \v = -c\, \v$,
where the constant~$c$ depends on the detailed form of the regularization parameters
in~\cite[Chapter~3]{cfs}). This is the usual equation for the Maxwell propagator. It involves 
the freedom in choosing a gauge.
For example, in the Lorenz gauge, one may choose~$S$ as the multiplication operator
in momentum space~$S(k) = c/k^2$. But~$S$ can also be given in any other gauge.
More generally, the choice of the Green's operator~\eqref{Sboson} always involves a choice of gauge.

\subsection{Discussion and Remarks} \label{secremark}
We now clarify the previous constructions by a few remarks.
We first note that, in order to simplify the computations, it is often convenient to assume that the
rescaling term in~\eqref{varyPsi} vanishes, i.e.
\beq \label{trcond}
\tr \big( \Psi(x)^*\, \delta \Psi(x)\big) = 0 \qquad \text{for all~$x \in M$}\:.
\eeq
This can be arranged for example by the transformation
\[ \delta \Psi \rightarrow \delta \Psi + \tr \big( \Psi(x)^*\, \delta \Psi(x)\big) \: \frac{\Psi}{c} \:. \]
Thinking in terms of the charts constructed in Section~\ref{secweo}, 
with the condition~\eqref{trcond} one restricts attention to a
special class of charts around~$x$.

We next point out that, as explained in~\cite[Section~2.5]{cfs}, the rescaling terms
in~\eqref{varyPsi} give rise to terms of higher order in~$\varepsilon/\l_\text{macro}$.
With this in mind, in many applications it is admissible to simply leave out the rescaling
and to ignore the condition~\eqref{trcond}.

We also remark that all the above jet spaces have a natural {\em{complex structure}}.
In order to understand how this comes about, we recall that according to~\eqref{varyPsi}
the vector fields on~$M$ were described by variations of the wave evaluation operator~\eqref{deltaPsi}.
Since the spin spaces are complex vector spaces, pointwise multiplication by complex scalars
gives a natural complex structure on~$\delta \Psi$. Using the notation~\eqref{varyPsi}, we thus obtain a
corresponding almost complex structure~$J$ on~$T_x\F$ given by
\beq \label{almostcomplex}
\begin{split}
&J\, \delta \hat{F}[\delta \Psi](x) = \delta \hat{F}[i \delta \Psi](x) \\
&= i \delta \Psi(x)^*\, \Psi(x) - i \Psi(x)^*\, \delta \Psi(x)
+ \frac{x}{c}\: \tr \big( -i \delta \Psi(x)^*\, \Psi(x) + i \Psi(x)^*\, \delta \Psi(x)\big) \:.
\end{split}
\eeq
This also gives rise to a complex structure on the vector spaces of vectorial jets on~$M$
like~$\Jfermi$ and~$\Jbose$. This complex structure is of no relevance for the
constructions in~\cite{cfs} but might be of importance for future developments.
Indeed, in~\cite{fockbosonic} an almost-complex structure was constructed on the jet spaces
in the more general setting of causal variational principles. It was used to
deduce a unitary time evolution on bosonic Fock spaces.
The detailed connection to the almost-complex structure in~\eqref{almostcomplex}
and generalization to fermionic Fock spaces still need to be worked out.

We finally point out that the choice of test jets in~\eqref{Jtestcl} is very restrictive.
In other words, the analysis of the continuum limit only uses very little of the information
contained in the EL equations. On the other hand, this information seems to capture
precisely what is needed in order to describe the effective macroscopic interaction.
One shortcoming of the analysis in the continuum limit is that the test jets
do not contain the bosonic jets,
\[ \Jtest \cap \Jbose = \varnothing \:. \]
This implies that the symplectic form as introduced in~\cite{jet} is undefined for the
bosonic jets. Moreover, since the intersection of the test jets with the fermionic jets
only contains the very restrictive class of jets formed of ultrarelativistic wave packets,
also the conserved surface layer integrals in~\cite{osi} cannot be evaluated
for interesting fermionic jets.
This last shortcoming is closely related to the fact that the Green's operator~\eqref{Sboson}
is purely bosonic, whereas the fermionic dynamics is encoded
in the Dirac equation~\eqref{s:diracPaux}.
Taking into account that in~\cite[Section~3.10]{cfs} the validity of the Dirac equation
is justified from the causal action principle by ruling out nonlocal potentials,
this procedure is conceptually convincing as a first step.
But eventually, one would like to have more general test jets, giving rise to a
unified description of the interaction in terms of Green's operators
composed of a fermionic and a bosonic component.
A first step in this direction is the computation of surface layer integrals 
for bosonic and fermionic jets in~\cite{action}.

\section{Example: Perturbation Expansion with Microscopic Mixing} \label{secmixing}
\subsection{Preliminaries}
The method of microscopic mixing of wave functions was introduced in~\cite{qft}
(based on preliminary considerations in~\cite{entangle}). 
Using our present notation, the basic construction is summarized as follows.
One first decomposes space-time into disjoint subsystems~$M_1, \ldots, M_L$,
\[ M = M_1 \cup \cdots \cup M_L \qquad \text{and} \qquad M_\as \cap M_\bs = \varnothing \quad
\text{if $a \neq b$}\:. \]
For each subsystem, one introduces a unitary operator~$V_\as$ with the property
that~$\1-V_\as$ is an operator of finite rank which maps particle and anti-particle states to
sea states and vice versa (for details see~\cite[Section~2.2]{qft}).
Then the kernel of the fermionic projector with microscopic mixing is introduced by
\begin{align}
P^\varepsilon(x,y) &= \sum_{\as, \bs =1}^L \chi_{M_\as}(x)
\: P^{\as, \bs}(x,y) \:\chi_{M_\bs}(y) 
\label{Pchar} \\
P^{\as, \bs}(x,y) &= -\Psi(x) \, V_\as  \,V_\bs^* \, \Psi(y)^* \label{Pmixansatz}
\end{align}
(where~$\chi_{M_\as}$ is the characteristic function).
In~\cite{qft}, this kernel of the fermionic projector is used as the starting point
for a perturbative treatment based on the methods of the analysis in the continuum limit.
It is shown that in a suitable limiting case, one obtains an effective interaction in terms of
bosonic and fermionic field operators acting on Fock spaces.

\subsection{A Synchronization Mechanism} \label{secsync}
In preparation for getting a connection to the setting of Section~\ref{secfragment},
we recast microscopic mixing in terms of the universal measure
(for a similar construction see~\cite[\S1.5.3]{cfs}). To this end, for a unitary operator~$V \in \U(\H)$
we define the measure~$V(\rho)$ by
\beq \label{Vrho}
(V \rho)(\Omega) = \rho \big(V \Omega V^{-1} \big) \:.
\eeq
We introduce the measure~$\hat{\rho}$ as the convex combination
\[ \hat{\rho} = \frac{1}{L} \sum_{\as=1}^L \rho_\as \qquad \text{with} \quad \rho_\as = V_\as \rho \:. \]
Then the resulting space-time~$\hat{M} := \supp \hat{\rho}$ is given by
\[ \hat{M} = \bigcup_{\as=1}^L M_\as \qquad \text{with} \qquad
M_\as := V_\as \,M\, V_\as^{-1} \:. \]
Comparing the unitary transformation~$x \rightarrow V x V^{-1}$ in~\eqref{Vrho}
with the first equation in~\eqref{xPid}, one sees that the wave evaluation operator~\eqref{weo}
is transformed to
\[ \hat{\Psi} \::\: \H \rightarrow C^0 \big(\hat{M}, S\hat{M} \big)\:,\qquad
\hat{\Psi}(x_\as) = \Psi(x) \, V_\as \:. \]
Applying this relation in the second equation in~\eqref{xPid}, one recovers~\eqref{Pmixansatz}.

Next, we rewrite the wave evaluation operator of the $\as^\text{th}$ subsystem as
\[ \hat{\Psi}_\as = \Psi + \Delta \Psi_\as \qquad \text{with} \qquad \Delta \Psi_\as = \Psi(x) \, \big(V_\as -\1 \big) \:. \]
Exactly as explained in Section~\ref{seccfsvacpert}, the resulting transformation of the universal
measure can be written as (cf.~\eqref{hatPsi} and~\eqref{hatrhoH})
\[ \rho_\as = (H_\as)_* \rho \:. \]
Expanding~$H_\as$ in a given chart on~$\F$ similar to~\eqref{wpex}, one obtains
inhomogeneities~$\v_\as^{(p)}$ in the EL equations which depend on the subsystem.
Following the constructions in Section~\ref{seclinfluct} for the linearized inhomogeneity~$\v=\v^{(1)}$,
one gets a corresponding linearized solution of the field equations~$\w^{(1)}$
which involves fluctuations. The higher orders in perturbation theory
are obtained just as in Section~\ref{secpertfluct}.
The crucial condition for the construction to work is that the
resulting fragmentation must be well-posed (see Definition~\ref{defwellposed}).

\begin{figure}
\begin{center}
{
\begin{pspicture}(-3,-0.9794922)(7.419669,0.9794922)
\psbezier[linecolor=black, linewidth=0.04](1.2899469,0.071167536)(0.9893741,-0.19104606)(1.12319,0.15230647)(0.9499469,0.186167534722224)(0.7767038,0.2200286)(0.8899469,-0.123832464)(0.7249469,-0.08883247)(0.5599469,-0.053832464)(0.6799469,0.39116752)(0.2799469,0.071167536)
\psdots[linecolor=black, dotsize=0.14375](1.3496691,0.08977865)
\rput[bl](0.31966913,-0.9035547){\normalsize{$\as$}}
\psline[linecolor=black, linewidth=0.08](0.012446899,0.08366753)(0.2874469,0.08366753)
\rput[bl](0.5096691,0.3014453){\normalsize{$S_\Fluct$}}
\psline[linecolor=black, linewidth=0.03](0.1574469,0.7136675)(0.1574469,-0.70133245)
\psline[linecolor=black, linewidth=0.03](1.3474469,0.67866755)(1.3474469,-0.7363325)
\psline[linecolor=black, linewidth=0.03](1.4874469,-0.48133245)(1.3474469,-0.7313325)(1.2074469,-0.48133245)
\psline[linecolor=black, linewidth=0.03](0.0174469,0.46866754)(0.1574469,0.7186675)(0.2974469,0.46866754)
\psline[linecolor=black, linewidth=0.03](0.2974469,-0.45633247)(0.1574469,-0.70633245)(0.0174469,-0.45633247)
\rput[bl](1.5146692,-0.8835547){\normalsize{$\as$}}
\rput[bl](1.9696691,-0.098554686){\normalsize{$+$}}
\psline[linecolor=black, linewidth=0.03](2.757447,0.67866755)(2.757447,-0.7363325)
\psdots[linecolor=black, dotsize=0.14](3.934669,0.084778644)
\psbezier[linecolor=black, linewidth=0.04](3.8949468,0.071167536)(3.5943742,-0.19104606)(3.72819,0.15230647)(3.554947,0.186167534722224)(3.3817039,0.2200286)(3.494947,-0.123832464)(3.329947,-0.08883247)(3.1649468,-0.053832464)(3.284947,0.39116752)(2.8849468,0.071167536)
\psline[linecolor=black, linewidth=0.08](2.622447,0.08366753)(2.8974469,0.08366753)
\psline[linecolor=black, linewidth=0.03](3.922447,0.67866755)(3.922447,-0.7363325)
\psline[linecolor=black, linewidth=0.03](2.617447,0.46866754)(2.757447,0.7186675)(2.8974469,0.46866754)
\psline[linecolor=black, linewidth=0.03](4.062447,-0.51133245)(3.922447,-0.76133245)(3.7824469,-0.51133245)
\psline[linecolor=black, linewidth=0.03](2.8974469,-0.49133247)(2.757447,-0.7413325)(2.617447,-0.49133247)
\rput[bl](2.914669,0.7214453){\normalsize{$\as$}}
\rput[bl](4.099669,-0.9035547){\normalsize{$\as$}}
\rput[bl](3.1896691,0.32144532){\normalsize{$S_\Fluct$}}
\end{pspicture}
}
\end{center}
\caption{Synchronization of fluctuations.}
\label{figsynchron}
\end{figure}
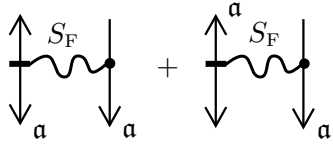

Choosing the jet spaces as in the continuum limit in Section~\ref{secchoosejet},
the above construction simplifies because the jet spaces do not have a scalar component.
In this limiting case, one recovers the perturbation expansion in~\cite{qft}
with one important exception: the perturbation expansion with fragmentation gives rise
to an additional {\em{synchronization mechanism}}.
Indeed, according to Definition~\ref{defbarS}, the Green's operators~$S_\Fluct$
acts on each subsystem separately,
\beq \label{Sfluct}
\big(S_\Fluct \big)^\as_\bs = \delta^\as_\bs\, S_\Fluct \:.
\eeq
From~\eqref{deltaFdef} one sees that it couples
only to the current generated by Dirac wave functions in the subsystem~$\as$
(see the Feynman diagrams in Figure~\ref{figsynchron}).
This seems to make it unnecessary to consider
the stochastic background field in~\cite[Section~4]{qft} for synchronization.
Also, the recombination of subsystems in~\cite[Section~7]{qft} needs to be reconsidered.
The consequences of this synchronization mechanism will be analyzed in detail
in a separate publication~\cite{qftlimit}.

\subsection{Gauge Potentials are Subsystem-Diagonal} 
The previous constructions
yield an interaction described by a Dirac equation which,
according to~\eqref{Sfluct}, is coupled to an electromagnetic potential
for each subsystem, i.e.
\[ (i \Pdd + \slashed{A}_\as - mY)\, \psi^\as(x) = 0 \qquad \text{for all~$\as =1,\ldots, L$}\:. \]
More generally, one could consider a matrix potential which mixes the subsystems, i.e.
\beq \label{Basbs}
\sum_{\bs=1}^L (i \Pdd + \slashed{A}^\as_\bs - mY)\, \psi^\bs = 0 \qquad \text{for all~$\as =1,\ldots, L$}\:.
\eeq
We now give an independent general argument which conveys a good intuitive understanding
for why such subsystem-mixing potentials must not occur.

The matrix potential in~\eqref{Basbs} can be regarded as a ~$\U(L)$ gauge potential.
To leading degree on the light cone, this gauge potentials affects the
kernel of the fermionic projector via generalized phase transformations
(for details see~\cite{light} or~\cite[\S3.6.2 and~\S4.3.2]{cfs}).
Considering for simplicity the special case of a gauge transformation, the Dirac wave functions
transforms according to
\[ \psi_\as(x) \rightarrow \sum_{\bs=1}^L U^\as_\bs(x) \, \psi_\bs(x) \:. \]
Using this transformation law in~\eqref{Pmixansatz} in the special case
with trivial mixing matrices~$V_1 = \cdots = V_L=\1$, one finds that the
kernel of the fermionic projector transforms according to
\[ P^{\as,\bs}(x,y) \rightarrow (U(x) \,v)^\as \: P(x,y)\: \overline{(U(y) \,v)^\bs} \:, \]
where
\[ v = (1,\ldots, 1) \in \C^L \:. \]
Since the Lagrangian is homogeneous of degree four in~$P(x,y)$, it transforms like
\[ \L(x,y) \rightarrow \sum_{\as,\bs=1}^L \big|(U(x) \,v)^\as\big|^4 \, \big|(U(y) \,v)^\bs\big|^4 \: \L(x,y)\:. \]
Thus, seeking for minimizers of the causal action, one must
\beq \label{varp}
\text{minimize} \qquad \sum_{\as=1}^L \big| (U v)^\as \big|^4 \:.
\eeq

We would like to show that the
minimizers of this functional are precisely the subsystem-diagonal potentials.
However, the situation is not quite so simple, as the following counter example shows:

\begin{Example} {\em{ Choose~$L=2$ and consider the one-parameter group of unitary
matrices~$(U_t)_{t \in \R}$
\[ U_t = \exp \left( \frac{i t}{2} \begin{pmatrix} 1 & 1 \\ 1 & 1 \end{pmatrix} \right) . \]
Using that the matrix in the exponent is twice a projection operator, a short computation yields
\[ U_t = \frac{1}{2} \begin{pmatrix} 1 & -1 \\ -1 & 1 \end{pmatrix}
+ \frac{e^{it}}{2} \begin{pmatrix} 1 & 1 \\ 1 & 1 \end{pmatrix} \:. \]
Thus
\[ U_t \,v = e^{it}\:  \begin{pmatrix} 1 \\ 1 \end{pmatrix} \]
Hence the relations
\[ \big| (U_t \,v)^\as \big| = 1 \qquad \text{for all $\as=1,2$} \]
hold, although the unitary operators~$U_t$ are not diagonal.
This shows that the diagonal unitary matrices cannot be singled out by minimizing~\eqref{varp}.
\QEDrem }}
\end{Example}

We now enter the general analysis. Given a compact connected Lie subgroup~$\G \subset \U(L)$, we set
\[ \G v := \{U v \:|\: U \in \G \} \subset \C^L \:. \]
Moreover, we introduce the {\em{diagonal}}
and {\em{orthogonal subgroups}} by
\begin{align}
\G^\text{\rm{d}} &= \Big\{ U \in \G \:\Big|\: U = \big( e^{i \varphi_1}, \ldots, e^{i \varphi_L} \big)
\text{ with } \varphi_\as \in \R \Big\} \label{Gd} \\
\G^\perp &= \Big\{ U \in \G \:\Big|\: U|_{\G v} = \1_{\G v} \Big\} \:. \label{Gortho}
\end{align}
The vector~$v$ is called {\em{cyclic}} if~$\G v = \C^L$.
Clearly, if~$v$ is cyclic, then~$\G^\perp$ is trivial.

\begin{Prp} \label{prpdigaonal}
The infimum of the functional in~\eqref{varp} is given by
\beq \label{infform}
\inf_{U \in \G} \sum_{\as=1}^L \big| (U v)^\as \big|^4 = \inf_{U \in \U(L)} \sum_{\as=1}^L \big| (U v)^\as \big|^4 = L \:.
\eeq
Moreover, if this functional is minimal on all of~$\G$, i.e.\
\[ \sum_{\as=1^L} \big| (Uv)^\as \big|^4 = L \qquad \text{for all~$U \in \G$} \:, \]
then every~$U \in \G$ has a unique decomposition into a diagonal and an orthogonal element,
\beq \label{Udec}
U = U^\text{\rm{d}} \,U^\perp \qquad \text{with} \qquad U^\text{\rm{d}} \in \G^\text{\rm{d}} \text{ and }
U^\perp \in \G^\perp \:.
\eeq
\end{Prp}
Before giving the proof, we explain what this result means.
Generally speaking, this proposition gives strong constraints for the form of the
subsystem-mixing gauge potentials.
Indeed, such potentials may be nontrivial only if the vector~$v$
is not cyclic. But the vector $v$ will be cyclic whenever each subsystem has its own dynamics.
Namely, in this case, the subsystem-diagonal gauge potentials will be different in each
subsystems, giving rise to different $\U(1)$-phases in each subsystem. As a consequence,
the group~$\G$ will contain the abelian subgroup of all diagonal unitary matrices,
implying that~$v$ is cyclic.

\Proof[Proof of Proposition~\ref{prpdigaonal}] We first prove~\eqref{infform}.
Since the rows of a unitary matrix are unit vectors, we know that
\[ \sum_{\bs=1}^L | U^\as_\bs |^2 =1\:. \]
As a consequence, using the Schwarz inequality,
\beq \label{schwarz}
L = \sum_{\as,\bs=1}^L \big| U^\as_\bs \big|^2
= \sum_{\as=1}^L \big| (Uv)^\as \big|^2 \leq \sqrt{L} \:\bigg( \sum_{\as=1}^L \big| (Uv)^\as \big|^4 \bigg)^\frac{1}{2} \:,
\eeq
implying that
\[  \sum_{\as=1}^L \big| (Uv)^\as \big|^4 \geq L \:. \]
Equality is attained in the case~$U=\1$, proving~\eqref{infform}.
More generally, equality holds if and only if all the summands in~\eqref{schwarz} coincide, i.e.
\beq \label{Uvrel}
\big| (Uv)^\as \big| = 1 \qquad \text{for all~$\as=1,\ldots, L$}\:.
\eeq

Next, we prove uniqueness of the decomposition~\eqref{Udec}. Suppose that
a unitary operator~$U$ has the representation~\eqref{Udec}.
Then, using~\eqref{Gortho}, we know that~$Uv = U^\text{\rm{d}} v$.
This relation uniquely determines all the phases~$\varphi_1, \ldots, \varphi_L$ in~\eqref{Gd}.
Hence~$U^\text{\rm{d}}$ is unique, which also determines~$U^\perp$ uniquely
by~$U^\perp = (U^\text{\rm{d}})^{-1} U$.

It remains to construct the decomposition~\eqref{Udec}.
Let~$A \in \g \subset \u(L)$ be a vector of the Lie algebra of~$\G$.
Then~\eqref{Uvrel} implies that for any vector~$w \in \G v$, the equation
\[ \big| (e^{i t A} w)^\as \big| = 1 \qquad \text{holds for all~$t \in \R$ and all~$\as=1,\ldots, L$}\:. \]
Employing a spectral decomposition of the Hermitian matrix~$A$,
\[ A = \sum_{k=1}^K \lambda_k \,E_k \:,\qquad e^{i t A} = \sum_{k=1}^K e^{i \lambda_k t} \,E_k \:, \]
we obtain
\beq \label{1rel}
1 = \big| (e^{i t A} w)^\as \big|^2 = \sum_{k,k'=1}^K e^{i(\lambda_k-\lambda_{k'})t}\:
\overline{\big( E_{k'} w \big)^\as}  \big(E_k w \big)^\as \:.
\eeq
We want to conclude that at most one summand is non-zero, i.e.
\[ \big(E_k w \big)^\as = 0 \qquad \text{for all~$k \neq \ell$} \]
and a suitable~$\ell = \ell(\as,w)$. To this end, assume conversely that~$(E_k w)^\as$
and~$(E_{k'} w)^\as$ are both non-zero for~$k \neq k'$. We choose~$k$ and~$k'$
such that~$\lambda_k - \lambda_{k'}$ is maximal. Then the right side of~\eqref{1rel} involves
non-zero Fourier terms $\sim e^{\pm i (\lambda_k- \lambda_k') t}$, a contradiction.

Let us show that~$\ell$ can be chosen independent of~$w$.
We proceed indirectly and assume that~$k:=\ell(\as,w_1) \neq \ell(\as,w_2)=:k'$.
Then, evaluating~\eqref{1rel} for~$w=w_1+w_2$, one gets a non-zero contribution
\[ \re \bigg( e^{i(\lambda_k-\lambda_{k'})t}\:
\overline{\big( E_{k'} w_2 \big)^\as}  \big(E_k w_1 \big)^\as \bigg) \:. \]
Varying the phase of~$w_2$, one again gets a contradiction.
We conclude that
\beq \label{f1}
\big(E_k w \big)^\as = 0 \qquad \text{for all~$k \neq \ell(\as)$ and all~$w \in \G v$}\:.
\eeq

Using the completeness of the spectral projectors, we obtain
\[ w^\as = \sum_{j=1}^L \big(E_j w \big)^\as = \sum_{j=\ell(\as)} \big(E_j w \big)^\as = \big(E_{\ell(\as)} w \big)^\as \:. \]
Combining this relation with~\eqref{f1}, it follows that
\[ (E_k w)^\as = \delta_{k, \ell(\as)}\: w^\as \:. \]
Since~$w \in \G v$ is arbitrary, we can also write this relation as
\[ E_k \big|_{\G v} = E^\text{\rm{d}}_k \big|_{\G v} \qquad \text{with} \qquad
\big(E^\text{\rm{d}}_k\big)^\as_\bs = \delta^\as_\bs\, \delta_{k, \ell(\as)}\:. \]
As a consequence, the matrix
\[ U^\text{\rm{d}} := \sum_{k=1}^K e^{i \lambda_k t} \, E^\text{\rm{d}}_k \]
is diagonal. Moreover, the matrix~$U^\perp := (U^\text{\rm{d}})^{-1} U$ is trivial on~$\G v$,
giving the desired decomposition~\eqref{Udec}.
\QED

\Thanks {{\em{Acknowledgments:}}
I would like to thank Johannes Kleiner and Sebastian Kindermann
for helpful discussions and suggestions on the manuscript.

\providecommand{\bysame}{\leavevmode\hbox to3em{\hrulefill}\thinspace}
\providecommand{\MR}{\relax\ifhmode\unskip\space\fi MR }
\providecommand{\MRhref}[2]{%
  \href{http://www.ams.org/mathscinet-getitem?mr=#1}{#2}
}
\providecommand{\href}[2]{#2}

\end{document}